\documentclass{aa}
\usepackage{natbib}
\usepackage{txfonts}
\usepackage{graphicx}
\bibpunct{(}{)}{;}{a}{}{,}

\def\cc{\rm ^{12}C}
\def\cz{\rm ^{13}C}
\def\nn{\rm ^{14}N}
\def\nz{\rm ^{15}N}

\begin{document}

\title{Combined model for $\nz$, $\cz$, and spin-state chemistry in molecular clouds}
\author{O. Sipil\"a\inst{1},
           L. Colzi\inst{2},
           E. Roueff\inst{3},
           P. Caselli\inst{1},
           F. Fontani\inst{4,3},
           and E. Wirstr\"om\inst{5}
}
\institute{Max-Planck-Institut f\"ur Extraterrestrische Physik (MPE), Giessenbachstr. 1, D-85748 Garching, Germany\\
e-mail: \texttt{osipila@mpe.mpg.de}
\and{Centro de Astrobiolog\'ia (CAB), CSIC-INTA, Ctra. de Ajalvir Km. 4, 28850 Torrej\'on de Ardoz, Madrid, Spain}
\and{Sorbonne Universit\'e, Observatoire de Paris, PSL University, CNRS, LERMA, 5 place Janssen, 92190 Meudon Cedex, France}
\and{INAF -- Osservatorio Astrofisico di Arcetri, Largo E. Fermi 5, 50125 Florence, Italy}
\and{Department of Space, Earth and Environment, Chalmers University of Technology, Onsala Space Observatory, 43992 Onsala, Sweden}
}

\date{Received / Accepted}

\abstract{We present a new gas-grain chemical model for the combined isotopic fractionation of carbon and nitrogen in molecular clouds. To this end, we have developed gas-phase and grain-surface chemical networks where the isotope chemistry of carbon and nitrogen is coupled with a time-dependent description of spin-state chemistry, which is important for nitrogen chemistry at low temperatures. We updated the rate coefficients of some isotopic exchange reactions considered previously in the literature, and here we present a set of new exchange reactions involving molecules substituted in $\cz$ and $\nz$ simultaneously. We applied the model to a series of zero-dimensional simulations representing a set of physical conditions across a prototypical prestellar core, exploring the deviations of the isotopic abundance ratios in the various molecules from the elemental isotopic ratios as a function of physical conditions and time. We find that the $\cc/\cz$ ratio can deviate from the elemental ratio to a high factor depending on the molecule, and that there are highly time-dependent variations in the ratios. The $\rm HCN/H\cz N$ ratio, for example, can obtain values of less than ten depending on the simulation time. The $\nn/\nz$ ratios tend to remain close to the assumed elemental ratio within approximately ten percent, with no clearly discernible trends for the various species as a function of the physical conditions. Abundance ratios between $\cz$-containing molecules and $\cz$+$\nz$-containing molecules however show somewhat increased levels of fractionation as a result of the newly included exchange reactions, though they still remain within a few tens of percent of the elemental $\nn/\nz$ ratio. Our results imply the existence of gradients in isotopic abundance ratios across prestellar cores, suggesting that detailed simulations are required to interpret observations of isotopically substituted molecules correctly, especially given that the various isotopic forms of a given molecule do not necessarily trace the same gas layers.}

\authorrunning{O. Sipil\"a et al.}
\titlerunning{Combined model for $\nz$, $\cz$, and spin-state chemistry}
\maketitle

\section{Introduction}

Isotopically substituted molecules are important probes of the physical and chemical conditions in the interstellar medium (ISM), especially in the shielded interior regions of molecular clouds where optical thickness effects become important and many common tracer species cannot be used. Furthermore, there is an increasing amount of observational evidence indicating that material in the Solar System has inherited a significant amount of chemical diversity from the natal cloud \citep[e.g.,][]{Altwegg19, Hily-Blant19, Drozdovskaya21, Grewal21}. Studies of isotopically substituted species can thus be used to investigate the connections between star-forming regions, exoplanets, and the Solar System \citep{Caselli12b, Ceccarelli14, Nomura22}.

Owing to various observational challenges, for example with the optical depth or the inability to access all isotopic forms of a given species simultaneously, observers often derive the $\nn/\nz$ ratio of a given species by using the so-called double isotope method, where an observed $\cz$/$\nz$ ratio is scaled by the elemental $\cc/\cz$ ratio \citep[e.g.,][]{Wampfler14,Colzi18b}. This implies the assumption that the isotopic abundance ratio in the studied molecule follows the elemental abundance ratio. \citet{Roueff15} have shown via chemical simulations that this assumption is generally not true, and indeed deviations of a factor of several from the elemental $\cc/\cz$ ratio are possible, depending on the simulation time. \citeauthor{Colzi20}\,(\citeyear{Colzi20}; hereafter C20) have recently come to the same conclusions, using a model with an updated list of exchange reactions. These results imply direct consequences for observationally deduced $\nn/\nz$ abundance ratios. We also note that \citet{loison2020} have recently published simulation results on carbon fractionation that are qualitatively similar to those of \citet{Roueff15} and C20.

Some $\nn/\nz$ abundance ratios can however be observed directly, and chemical models have struggled to reproduce observed $\nz$ fractionation even after the inclusion of a variety of isotopic exchange reactions \citep[e.g.,][]{Roueff15, Wirstrom18, loison2019}. A particular example is the $\rm N_2H^+/ \nz NH^+$ ratio, which presents a high degree of anti-fractionation (ratios higher than the elemental value, with an excess of the $\nn$ containing species with respect to the $\nz$ ones) in observations of prestellar cores \citep{Bizzocchi13, Redaelli18}, while models usually predict fractionation. It has been suggested that the discrepancy might arise as a result of unphysically large variations in the isotope-dependent dissociative recombination rate of $\rm N_2H^+$ \citep{loison2019,Hily-Blant20}, or owing to isotope-selective photodissociation \citep{Heays14,Visser18, Furuya18, Colzi19}. The latter option is less controversial, but even the models of \citet{Furuya18} with the selective photodissociation included cannot reproduce the $\nz$ anti-fractionation in $\rm N_2H^+$. The (anti-)fractionation trends seem to depend on species though, and indeed the $\nn/\nz$ ratio in ammonia has been recently observed to lie close to the elemental ratio \citep{Redaelli23}, which is in agreement with the earlier simulation predictions.

One effect that has, to our knowledge, not been explored earlier in isotopic fractionation models is exchange reactions that lead to multiple fractionation, that is, allowing atom exchanges between species that may contain simultaneous isotopic substitution in multiple elements. In this paper, we present the first model that includes this effect; our model connects carbon and nitrogen isotopic fractionation and also includes a description of spin-state chemistry for light hydrogen-containing species. We employed the new model to study (combined) carbon and nitrogen fractionation in physical conditions representing a region in and around a prestellar core, searching especially for signs of gradients in the spatial distributions of isotopic abundance ratios that have been recently inferred observationally toward the prestellar core L1544 \citep{Spezzano22b}. We also included isotope-selective photodissociation of $\rm N_2$ and studied its effect on the abundances of nitrogen-bearing species, and we paid special attention to deviations of the simulated isotopic abundance ratios from the corresponding elemental ratios.

The present work is theoretical in nature, and we focus on the presentation of the newly developed chemical model. An in-depth comparison of the results of the model to various observations, also including objects and environments other than prestellar cores, will be carried out in a future follow-up work.

The paper is structured as follows. In Sect.\,\ref{s:model}, we introduce the model and present lists of the exchange reactions included in this work. We present the chemical simulation results in Sect.\,\ref{s:results}, and discuss their implications in Sect.\,\ref{s:discussion}. We give our concluding remarks in Sect.\,\ref{s:conclusions}.

\section{Description of the model}\label{s:model}

\subsection{Chemical network}\label{ss:chemicalNetwork}

We have developed a new model to describe the isotopic fractionation of $\nz$ and $\cz$ in molecular clouds. The model includes exchange reactions involving both of these elements separately, as well as reactions leading to multiple fractionation, that is, forming species with simultaneous $\nz$ and $\cz$ substitution.

The chemical network is constructed following the same principle as presented in C20, and we refer the reader to Sect.\,2 of that paper for more details including the relevant references. There are three major differences in the construction of the chemical network in the present work compared to C20. First, as stated above, we allow for multiple fractionation, whereas in C20 only the fractionation of carbon was considered. Second, we introduce spin-state chemistry for $\rm H_2$, $\rm H_2^+$, and $\rm H_3^+$ following the approach described in \citet{Sipila15b}. This inclusion is important in the present context because of the sensitivity of nitrogen chemistry to the $\rm H_2$ ortho-to-para ratio \citep[e.g.,][]{Dislaire12,Wirstrom12,Roueff15} -- omitting it in C20 was however justified because carbon (isotope) chemistry does not depend on spin chemistry. Third, we track the positions of the isotopic forms across the molecules, so the model distinguishes between $\rm H\cz CCN$, $\rm HC\cz CN$, and $\rm HCC\cz N$, for example.

We employ an updated version of the chemical network creation procedure discussed in C20, in which the KIDA gas-phase network \citep{Wakelam15} is used as the base network on top of which isotope chemistry is added. For grain-surface chemistry we use our own custom network that is originally based on that of \citet{Semenov10}. The isotopes are introduced one by one, with the network creation proceeding in three steps. In the first step, we introduce $\cz$ along with the appropriate exchange reactions (see below). In the second step, $\nz$ is added on top of the network created in the first step. At the end of this step, exchange reactions are introduced for $\nz$ and for $\nz$+$\cz$. In the third step, spin-state chemistry is introduced for those reactions where $\rm H_2$, $\rm H_2^+$, or $\rm H_3^+$ appear. We intentionally limit the carbon isotope chemistry to molecules that contain up to three C atoms, and the nitrogen isotope chemistry to molecules that contain at most two N atoms. In this way we are able to keep the number of reactions in the resulting chemical network within manageable limits. For the same reason we do not consider deuterium chemistry, or spin chemistry for molecules containing elements heavier than hydrogen, though this is possible from a technical standpoint. With these restrictions, the total number of reactions in the model is approximately 35\,000, 1700 of which are grain-surface reactions.

The isotopic exchange reactions included in the model are collected in Tables~\ref{tab:reaccolzi}~and~\ref{tab:reaccolzi2}. These include a set of new exchange reactions that we propose for cases where $\rm ^{13}C$ and $\rm ^{15}N$ appear simultaneously. Their significance for the chemistry is assessed in Sect.\,\ref{ss:doubleSubstitution}. We note that the list given in Tables~\ref{tab:reaccolzi}~and~\ref{tab:reaccolzi2} covers exchange reactions that would be identity reactions when the isotopes are not distinguished (with the exception of $\rm N^+ + H_2 \rightarrow NH^+ + H$). Such reactions are not automatically produced as a result of the procedure with which we generate the fractionation network (see also C20), and so it is essential to add the reactions in Tables~\ref{tab:reaccolzi}~and~\ref{tab:reaccolzi2} on top of the automatically generated network to make sure that important reaction pathways are not missed. We note that the automated procedure does not take energy differences into account either.

\begin{table*}
\setlength{\tabcolsep}{2pt}
\begin{center}
\caption{List of carbon isotopic exchange reactions included in the model.}
\begin{tabular}{lclccc}
  \hline
  \hline
  Label  & Reaction & $k_{\rm f}$ & $f(B, m)$\tablefootmark{a} & $\Delta E$\tablefootmark{b}\\ 
          & & (cm$^{3}$ s$^{-1}$) & & (K)& Reference  \\
          \hline
          \multicolumn{6}{c}{C isotopic exchange reactions included in \citet{Colzi20}.}\\ 
  \hline
  (1) B & $^{13}$C$^{+}$ + CO $\rightleftharpoons$ $^{12}$C$^{+}$ + $^{13}$CO & 6.6$\times$10$^{-10}\times (\frac{T}{300})^{-0.45} \times$ &1 & 34.7  & (1)\\
& & $ \exp(-6.5/T) \times \frac{1}{1+\exp(-34.7/T)}$ & & \\
(2) A & $^{13}$CO + HCO$^{+}$ $\rightleftharpoons$ CO + H$^{13}$CO$^{+}$ & 2.6$\times$10$^{-10} \times (\frac{T}{300})^{-0.4}$ &1 & 17.4 & (1) \\
(3) B& $^{13}$C$^{+}$ + CN $\rightleftharpoons$ $^{12}$C$^{+}$ + $^{13}$CN & 3.82$\times$10$^{-9} \times  (\frac{T}{300})^{-0.4}\times \frac{1}{1+\exp(-31.1/T)}$  &1 & 31.1 & (1)\\
(4) B& $^{13}$C + CN $\rightleftharpoons$ $^{12}$C + $^{13}$CN & 3.0$\times$10$^{-10} \times \frac{1}{1+\exp(-31.1/T)}$ & 1 & 31.1 & (1) \\
(5) B& $^{13}$C + C$_{2}$ $\rightleftharpoons$ $^{12}$C + $^{13}$CC & 3.0$\times$10$^{-10} \times \frac{2}{2+\exp(-25.9/T)}$ & 2 & 25.9 & (1) \\
 (6) B& $^{13}$C$^{+}$ + C$_{2}$ $\rightleftharpoons$ $^{12}$C$^{+}$ + $^{13}$CC &1.86$\times$10$^{-09}\times \frac{2}{2+\exp(-25.9/T)}$ & 2 & 25.9 & (2)\\
 (7) B&  $^{13}$C$^{+}$ + $^{13}$CC  $\rightleftharpoons$ $^{12}$C$^{+}$ + $^{13}$C$_{2}$ & 1.86$\times$10$^{-09}\times \frac{0.5}{0.5+\exp(-26.4/T)}$ & 0.5 & 26.4 & (2)\\
 (8) B& $^{13}$C + $^{13}$CC $\rightleftharpoons$ $^{12}$C + $^{13}$C$_{2}$ & 3.0$\times$10$^{-10}\times \frac{0.5}{0.5+\exp(-26.4/T)}$ & 0.5 & 26.4 & (2)\\
 (9) B& $^{13}$C$^{+}$ +CS $\rightleftharpoons$ $^{12}$C$^{+}$ + $^{13}$CS & 1.86$\times$10$^{-09}\times \frac{1}{1+\exp(-26.3/T)}$  & 1 & 26.3 & (2)\\
 (10) B & $^{13}$C + C$_{3}$ $\rightleftharpoons$ $^{12}$C + $^{13}$CC$_{2}$ & 3.0$\times$10$^{-10}\times \frac{2}{2+\exp(-27/T)}$ & 2 & 27 & (2)\\
  (11) B& $^{13}$C$^{+}$ + C$_{3}$ $\rightleftharpoons$ $^{12}$C$^{+}$ + $^{13}$CC$_{2}$ & 1.8$\times$10$^{-9}\times \frac{2}{2+\exp(-27/T)}$  & 2 & 27 & (2)\\
 \hline
 \multicolumn{6}{c}{Newly included C isotopic exchange reactions. See also \citet{loison2020}.}\\ \hline
(12)B& $^{12}$C$^{+}$ + $^{13}$CCC $\rightleftharpoons$ $^{12}$C$^{+}$ + C$^{13}$CC & 1.8$\times$10$^{-9}\times \frac{0.5}{0.5+\exp(-16/T)}$  & 0.5 & 16 & (2), (3)\\
(13)B& $^{13}$C$^{+}$ + C$_{3}$ $\rightleftharpoons$ $^{12}$C$^{+}$ + C$^{13}$CC & 1.8$\times$10$^{-9}\times \frac{1}{1+\exp(-43/T)}$  & 1 & 43 & This work\\
(14) B & $^{12}$C + $^{13}$CCC $\rightleftharpoons$ $^{12}$C +  C$^{13}$CC & 3.0$\times$10$^{-10}\times \frac{0.5}{0.5+\exp(-16/T)}$ &  0.5 & 16 & (2), (3)\\
(15) B & $^{13}$C + C$_{3}$ $\rightleftharpoons$ $^{12}$C +  C$^{13}$CC & 3.0$\times$10$^{-10}\times \frac{1}{1+\exp(-43/T)}$ &  1 & 43 & This work\\
(16) B& $^{13}$C + $^{13}$CC$_{2}$ $\rightleftharpoons$ $^{12}$C + C$^{13}$C$_{2}$ & 3.0$\times$10$^{-10}\times \frac{1}{1+\exp(-43/T)}$  &  1 & 43 & This work. See also (3)\\
(17) B& $^{13}$C + $^{13}$CC$_{2}$ $\rightleftharpoons$ $^{12}$C + $^{13}$CC$^{13}$C & 3.0$\times$10$^{-10}\times \frac{0.5}{0.5+\exp(-27/T)}$  &  0.5 & 27 & This work. See also (3)\\
(18) B& $^{13}$C + C$^{13}$CC $\rightleftharpoons$ $^{12}$C + C$^{13}$C$_{2}$ & 3.0$\times$10$^{-10}\times \frac{2}{2+\exp(-27/T)}$  &  2 & 27 & This work. See also (3)\\
(19) B& $^{13}$C + C$^{13}$CC $\rightleftharpoons$ $^{12}$C + $^{13}$CC$^{13}$C & 3.0$\times$10$^{-10}\times \frac{1}{1+\exp(-11/T)}$  &  1 & 11 & This work. See also (3)\\
(20) B& $^{13}$C$^{+}$ + $^{13}$CC$_{2}$ $\rightleftharpoons$ $^{12}$C$^{+}$ + C$^{13}$C$_{2}$ & 1.8$\times$10$^{-9}\times \frac{1}{1+\exp(-43/T)}$  &  1 & 43 & This work. See also (3)\\
(21) B& $^{13}$C$^{+}$ + $^{13}$CC$_{2}$ $\rightleftharpoons$ $^{12}$C$^{+}$ + $^{13}$CC$^{13}$C & 1.8$\times$10$^{-9}\times \frac{0.5}{0.5+\exp(-27/T)}$  &  0.5 & 27 & This work. See also (3)\\
(22) B& $^{13}$C$^{+}$ + C$^{13}$CC $\rightleftharpoons$ $^{12}$C$^{+}$ + C$^{13}$C$_{2}$ & 1.8$\times$10$^{-9}\times \frac{2}{2+\exp(-27/T)}$  &  2 & 27 & This work. See also (3)\\
(23) B& $^{13}$C$^{+}$ + C$^{13}$CC $\rightleftharpoons$ $^{12}$C$^{+}$ + $^{13}$CC$^{13}$C & 1.8$\times$10$^{-9}\times \frac{1}{1+\exp(-11/T)}$  &  1 & 11 & This work. See also (3)\\

(24) B & $^{13}$C + HCN $\rightleftharpoons$ $^{12}$C +  H$^{13}$CN &2.0$\times$10$^{-10}\times \frac{1}{1+\exp(-48/T)}$ & 1 & 48 & (3) \\
(25) B & $^{13}$C + HNC $\rightleftharpoons$ $^{12}$C +  HN$^{13}$C &3.0$\times$10$^{-11}\times \frac{1}{1+\exp(-33/T)}$ & 1 & 33 & (3) \\
(26)B & $^{13}$C + HCNH$^{+}$ $\rightleftharpoons$ $^{12}$C +  H$^{13}$CNH$^{+}$ &1.0$\times$10$^{-9}\times \frac{1}{1+\exp(-50/T)}$ & 1 & 50 & (3) \\
(27) B& $^{13}$C +CS $\rightleftharpoons$ $^{12}$C + $^{13}$CS & 2.0$\times$10$^{-10}\times \frac{1}{1+\exp(-26.3/T)}$  & 1 & 26.3 & (3)\\
(28)  B& H+ $^{13}$CCH  $\rightleftharpoons$ H+ C$^{13}$CH  & 2.0$\times$10$^{-10}\times \frac{1}{1+\exp(-8.1/T)}$  & 1 & 8.1 & (3)\\
(29-a)  B& H+ c-$^{13}$CCCH$_{2}$  $\rightleftharpoons$ H+ c-C$^{13}$CCH$_{2}$  & 2.0$\times$10$^{-10}\times \frac{2}{2+\exp(-26/T)}$  & 2 & 26 & (3)\\
(29-b)  B& H+ c-$^{13}$CCCH$_{2}$  $\rightleftharpoons$ H+ c-CC$^{13}$CH$_{2}$   & 2.0$\times$10$^{-10}\times \frac{2}{2+\exp(-26/T)}$  & 2 & 26 & (3)\\\
(30)  B& H+ $^{13}$CCS  $\rightleftharpoons$ H+ C$^{13}$CS  & 4.0$\times$10$^{-11}\times \frac{1}{1+\exp(-18/T)}$  & 1 & 18 & (3)\\\
(31) B & HCNH$^{+}$ + H$^{13}$CN $\rightleftharpoons$ H$^{13}$CNH$^{+}$ + HCN & 2.0$\times$10$^{-9}\times (\frac{T}{300})^{-0.5} \times \frac{1}{1+\exp(-2.9/T)}$ & 1 & 2.9 & (3)\ \\
\hline

  \normalsize
  \label{tab:reaccolzi}
  \end{tabular}
  \end{center}
\tablefoot{Type~A reactions are direct reactions, while type~B reactions are those involving adduct formation, without isomerization, as defined by \citet{Roueff15}.\\
\tablefoottext{a}{$f(B, m)$ is a probability factor that depends on the rotational constant, mass, and symmetry factors of the reactants and products. In reactions involving $^{13}$C, the mass ratio of the reactants and the products is close to unity. Then, $f(B, m)= q({\rm C})q({\rm D})$/$q({\rm A})q({\rm B})$, where A and B are the reactants, C and D the products, and $q(...)$ are the internal molecular partition functions.}
\tablefoottext{b}{$\Delta E$ is the exoergicity of the exchange reaction in the forward direction.}
}
\tablebib{
(1)~\citet{Roueff15}; (2)~\citet{Colzi20}; (3)~\citet{loison2020}}
 \end{table*}

 \begin{table*}
 \setlength{\tabcolsep}{1.5pt}
\begin{center}
\caption{List of nitrogen, as well as carbon+nitrogen, isotopic exchange reactions included in the model.}
\begin{tabular}{lclccc}
  \hline
  \hline
  Label  & Reaction & $k_{\rm f}$ & $f(B, m)$\tablefootmark{a} & $\Delta E$\tablefootmark{b}\\ 
          & & (cm$^{3}$ s$^{-1}$) & & (K) & Reference \\
      \hline  
  \multicolumn{6}{c}{N isotopic exchange reactions from \citet{Roueff15}.}\\ 
  \hline
(32) A & N$^{15}$N + N$_{2}$H$^{+}$ $\rightleftharpoons$ N$^{15}$NH$^{+}$ + N$_{2}$ & 2.3$\times$10$^{-10}$ & 0.5 & 10.3 &(1), (2)\\
(33) A& N$^{15}$N + N$_{2}$H$^{+}$ $\rightleftharpoons$ $^{15}$NNH$^{+}$ + N$_{2}$ & 2.3$\times$10$^{-10}$ & 0.5 & 2.1 &(1), (2)\\
 (34) A& N$^{15}$N + $^{15}$NNH$^{+}$ $\rightleftharpoons$ N$^{15}$NH$^{+}$ + N$^{15}$N & 4.6$\times$10$^{-10}$ & 1 & 8.1 &(1), (2)\\
 (35) B& $^{15}$N$^{+}$ + N$_{2}$ $\rightleftharpoons$ $^{14}$N$^{+}$ + N$^{15}$N & 4.8$\times$10$^{-10} \times \frac{2}{2+\exp(-28.3/T)}$ &2& 28.3 &(1), (3)\\
 (36) C& $^{15}$N + CNC$^{+}$ $\rightleftharpoons$ C$^{15}$NC$^{+}$ + $^{14}$N & 3.8$\times$10$^{-12} \times (\frac{T}{300})^{-1}$ &1 & 38.1 & (1) \\
 (37) B& $^{15}$N + C$_{2}$N $\rightleftharpoons$ $^{14}$N + C$_{2}^{15}$N & 1.6$\times10^{-10}\times (\frac{T}{300})^{1/6} \times \frac{1}{1+\exp(-26.7/T)}$   &1 & 26.7 &(1) \\
  \hline
   \multicolumn{6}{c}{N isotopic exchange reactions from \citet{loison2019}.}\\ \hline
 (38) B & NH$_{4}^{+}$ + $^{15}$NH$_{3}$ $\rightleftharpoons$ $^{15}$NH$_{4}^{+}$ + NH$_{3}$ & 1.3$\times 10^{-9}\times (\frac{T}{300})^{-0.5} \times \frac{1}{1+\exp(-14.5/T)}$   &1 & 14.5  &(4)\\
 (39) B & HCNH$^{+}$ + HC$^{15}$N $\rightleftharpoons$HC$^{15}$NH$^{+}$ + HCN & 2.0$\times10^{-9}\times (\frac{T}{300})^{-0.5} \times \frac{1}{1+\exp(-10.1/T)}$  &1 & 10.1  &(4)\\
 \hline
  \multicolumn{6}{c}{Spin-state dependent $\rm NH^+$ formation reactions involving $^{14}$N and $^{15}$N.}\\ \hline
    (40) &$^{14}$N$^{+}$ + p-H$_{2}$ $\rightarrow$  $^{14}$NH$^{+}$ + H & 8.35$\times 10^{-10}\times \exp(-168.5/T)$  & -- & 168.5 & (1), (5)\\
  (41) &$^{14}$N$^{+}$ + o-H$_{2}$ $\rightarrow$  $^{14}$NH$^{+}$ + H & 4.2$\times10^{-10}\times (\frac{T}{300})^{-0.17}\times \exp(-44.5/T)$  & -- & 44.5 & (1), (5)\\
     (42) &$^{15}$N$^{+}$ + p-H$_{2}$ $\rightarrow$  $^{15}$NH$^{+}$ + H & 8.35$\times10^{-10}\times \exp(-164.3/T)$  & -- & 164.3 & (1)\\
 (43) &$^{15}$N$^{+}$ + o-H$_{2}$ $\rightarrow$  $^{15}$NH$^{+}$ + H & 4.2$\times10^{-10}\times (\frac{T}{300})^{-0.17}\times \exp(-39.7/T)$  & -- & 39.7 & (1)\\
    \hline  
  \multicolumn{6}{c}{New C+N isotopic exchange reactions proposed here.}\\ \hline 
  (44) B & $^{13}$C$^{+}$ + C$^{15}$N $\rightleftharpoons$  C$^{+}$ + $^{13}$C$^{15}$N & 3.8$\times$10$^{-9}\times (\frac{T}{300})^{-0.4} \times \frac{1}{1+\exp(-31.6/T)}$ & 1 & 31.6 & This work, (6)-(9)\\
   (45) B & $^{13}$C + C$^{15}$N $\rightleftharpoons$  C + $^{13}$C$^{15}$N & 3.0$\times$10$^{-10}\times  \frac{1}{1+\exp(-31.6/T)}$ & 1 & 31.6 & This work, (6)-(9) \\
   (46) B & $^{13}$C + HC$^{15}$N $\rightleftharpoons$ C + H$^{13}$C$^{15}$N & 2.0$\times$10$^{-10}\times  \frac{1}{1+\exp(-47.5/T)}$ & 1 & 47.5 & This work, (10)\\
   (47) B & $^{13}$C + H$^{15}$NC $\rightleftharpoons$ C + H$^{15}$N$^{13}$C & 3.0$\times$10$^{-11}\times  \frac{1}{1+\exp(-33.8/T)}$ & 1 & 33.8& This work, (10) \\
   (48) B & $^{13}$C + HC$^{15}$NH$^{+}$ $\rightleftharpoons$ C + H$^{13}$C$^{15}$NH$^{+}$ & 1.0$\times$10$^{-9}\times  \frac{1}{1+\exp(-50.8/T)}$ & 1 & 50.8 & This work, (10)\\
   (49) B & HC$^{15}$N + H$^{13}$CNH$^{+}$ $\rightleftharpoons$ H$^{13}$CN + HC$^{15}$NH$^{+}$ & 2.0$\times$10$^{-9}\times (\frac{T}{300})^{-0.5} \times  \frac{1}{1+\exp(-6.9/T)}$ & 1 & 6.9 & This work, (10) \\
    (50) B & H$^{13}$CN + HC$^{15}$NH$^{+}$ $\rightleftharpoons$ HCN + H$^{13}$C$^{15}$NH$^{+}$ & 2.0$\times$10$^{-9}\times (\frac{T}{300})^{-0.5} \times  \frac{1}{1+\exp(-2.4/T)}$ & 1 & 2.4& This work, (10) \\ 
    (51) B & HC$^{15}$N + H$^{13}$CNH$^{+}$ $\rightleftharpoons$ HCN + H$^{13}$C$^{15}$NH$^{+}$ & 2.0$\times$10$^{-9}\times (\frac{T}{300})^{-0.5} \times  \frac{1}{1+\exp(-10.5/T)}$ & 1 & 10.5 & This work, (10) \\
   \hline
   \normalsize
  \label{tab:reaccolzi2}
  \end{tabular}
  \end{center}
\tablefoot{Type~A reactions are direct reactions, type~B reactions involve adduct formation without isomerization, and type~C reactions involve adduct formation with possible isomerization pathways, as defined by \citet{Roueff15}.\\
\tablefoottext{a}{$f(B, m)$ is a probability factor that depends on the rotational constant, mass, and symmetry factors of the reactants and products. In reactions involving $^{13}$C, the mass ratio of the reactants and the products is close to unity. Then, $f(B, m)= q({\rm C})q({\rm D})$/$q({\rm A})q({\rm B})$, where A and B are the reactants, C and D the products, and $q(...)$ are the internal molecular partition functions.}
\tablefoottext{b}{$\Delta E$ is the exoergicity of the exchange reaction in the forward direction.}}
\tablebib{
(1)~\citet{Roueff15}; (2)~\citet{adams1981}; (3)~\citet{anicich1977}; (4)~\citet{loison2019}; (5)~\citet{Dislaire12}; (6)~\citet{Ram10}; (7)~\citet{Ram12}; (8)~\citet{Colin12}; (9)~\citet{Colin14}; (10)~\citet{Mehnen22}}
 \end{table*}
 
 \subsection{Simulation parameters}
 
The focus of the present paper is in the presentation of the newly developed isotope chemistry model. Therefore, unlike in C20, we do not attempt to reproduce observations of isotopic abundance ratios in the ISM, specifically in molecular clouds. Such an effort is reserved for an upcoming work (Colzi et al., in prep.), where we will also explore the effect of various parameters, like the cosmic ray ionization rate, on the isotopic ratios. Here, we run a set of zero-dimensional chemical simulations representing the typical physical conditions across a prestellar core and the surrounding envelope. Table~\ref{tab:simulationParameters} collects the parameters of the simulations, including details on the chemical network used in each simulation, and on the adopted physical parameters. In particular, in one simulation we test the effect of isotope-selective photodissociation for $\rm N_2$, considering that this has been invoked as an explanation for observed N fractionation gradients across the prestellar core L1544 \citep{Spezzano22b}. The simulations S1, S2, and~S3 represent, respectively, the center, inner region, and the outer envelope of a prototypical prestellar core. The values of the physical parameters not already given in Table~\ref{tab:simulationParameters} are kept the same as in C20, and are reproduced in Table~\ref{tab:physicalParameters} for the convenience of the reader. The initial elemental abundances are displayed in Table~\ref{tab:initialabundances}. We have chosen the elemental carbon and nitrogen isotopic ratios to correspond to the values adopted in previous chemical simulations (e.g., \citealt{Roueff15}; \citealt{loison2019}): 68 for carbon \citep{Milam05}, and 440 for nitrogen \citep{Marty11}. We choose a low fiducial value of $10^{-3}$ for the $\rm H_2$ ortho-to-para ratio (OPR), motivated by earlier simulation studies \citep[e.g.,][]{Faure19,Lupi21}. However, as the OPR is expected to influence nitrogen hydride chemistry, we have also tested its effect on the simulation results. These tests are described in Sect.\,\ref{ss:H2opr}.

\begin{table*}
\begin{center}
\caption{Parameters of the chemical simulations. }
\begin{tabular}{ccccc}
\hline
\hline
Simulation label & $n({\rm H_2}) \,[\rm cm^{-3}]$ & $T_{\rm gas}, T_{\rm dust} \,[\rm K]$ & $A_{\rm V} \, [\rm mag]$ & Chemical network \\
\hline
S1 & $10^6$ & 7, 7 & 30 & Full chemical network with C and N fractionation, and \\
                                                & & & & spin-state chemistry \\
S2 & $10^4$ & 10, 10 & 10 & As in S1 \\
S2\_3C & $10^4$ & 10, 10 & 10 & Only carbon fractionation, no spin-state chemistry \\
S2\_3C\_alt & $10^4$ & 10, 10 & 10 & As in S2\_3C, but omitting the newly included C isotopic \\
                                                & & & & exchange reactions displayed in Table~\ref{tab:reaccolzi} \\
S2\_C20 & $10^4$ & 10, 10 & 10 & Chemical networks of C20 \\
S3 & $5 \times 10^2$ & 20, 15 & 1 & As in S1  \\
S3\_ss & $5 \times 10^2$ & 20, 15 & 1 & As in S1, but with isotope-selective $\rm N_2$ and CO self-shielding$^{(a)}$ \\
\hline
\label{tab:simulationParameters}
\end{tabular}
\end{center}
\tablefoot{The cosmic ray ionization rate is fixed to $\zeta = 1.3 \times 10^{-17} \, \rm s^{-1}$ in all simulations. $^{(a)}$ The assumed values for the $\rm N_2$ and CO column densities are $\rm 10^{13} \, cm^{-2}$ and $\rm 9 \times 10^{14} \, cm^{-2}$, respectively (see Sect.\,\ref{ss:N2selfShielding}). $\rm N_2$ and CO self-shielding effects are neglected in all other simulations. $\rm H_2$ self-shielding is however taken into account in all simulations, with the column density calculated from the relation $N({\rm H_2})/A_{\rm V} = 10^{21} \, \rm cm^{-2} \, mag^{-1}$ and the self-shielding factor calculated following \citet{Draine96} (their equation~37).}
\end{table*}

\begin{table}
        \centering
        \caption{Values of the physical parameters kept fixed in all simulations.}
        \begin{tabular}{lc}
                \hline
                \hline
                Parameter & Value\\
                \hline
                cosmic ray ionization rate ($\zeta$) & $1.3\times10^{-17} \, \rm s^{-1}$\\
                grain radius ($a_{\rm g}$) & $10^{-5} \, \rm cm$\\
                grain material density ($\rho_{\rm g}$) & $3 \, \rm g \, \rm cm^{-3}$\\
                diffusion-to-binding energy ratio ($E_{\rm d}/E_{\rm b})$ & $0.6$\\
                dust-to-gas mass ratio ($R_{\rm d}$) & $0.01$\\
                \hline
        \end{tabular}
        \label{tab:physicalParameters}
\end{table}

\begin{table}
        \centering
        \caption{Initial elemental abundances (with respect to $n_{\rm H} \approx 2\,n({\rm H_2})$).}
        \begin{tabular}{cc}
                \hline
                \hline
                Species & Abundance\\
                \hline
                $\rm H_2$ & $5.00\times10^{-1}$\\
                $\rm He$ & $9.00\times10^{-2}$\\
                $\rm C^+$ & $1.20\times10^{-4}$\\
                $\rm ^{13}C^+$ & $1.76\times10^{-6}$\\
                $\rm N$ & $7.60\times10^{-5}$\\
                 $\rm ^{15}N$ & $1.72\times10^{-7}$\\
                $\rm O$ & $2.56\times10^{-4}$\\
                $\rm S^+$ & $8.00\times10^{-8}$\\
                $\rm Si^+$ & $8.00\times10^{-9}$\\
                $\rm Na^+$ & $2.00\times10^{-9}$\\
                $\rm Mg^+$ & $7.00\times10^{-9}$\\
                $\rm Fe^+$ & $3.00\times10^{-9}$\\
                $\rm P^+$ & $2.00\times10^{-10}$\\
                $\rm Cl^+$ & $1.00\times10^{-9}$\\
                $\rm F$ & $2.00\times10^{-9}$\\
                \hline
        \end{tabular}
        \label{tab:initialabundances}
        \tablefoot{The initial $\rm H_2$ OPR is $1 \times 10^{-3}$. The assumed $\rm ^{12}C/^{13}C$ and $\rm ^{14}N/^{15}N$ elemental ratios are 68 and 440, respectively.}
\end{table}

\section{Results}\label{s:results}

In what follows, we discuss the results of the chemical simulations for carbon and nitrogen isotopologues separately (Sects.\,\ref{ss:C20comparison} and \ref{ss:nitrogenIsotopologues}, respectively). In the former case, we emphasize the comparison of the present model to that of C20 given the significant differences in the construction of the model. We also present results for the combined fractionation of carbon and nitrogen (Sect.\,\ref{ss:C+N}).

\subsection{Carbon isotopologues; comparison to C20}\label{ss:C20comparison}

\begin{figure*}
\centering
        \includegraphics[width=2.0\columnwidth]{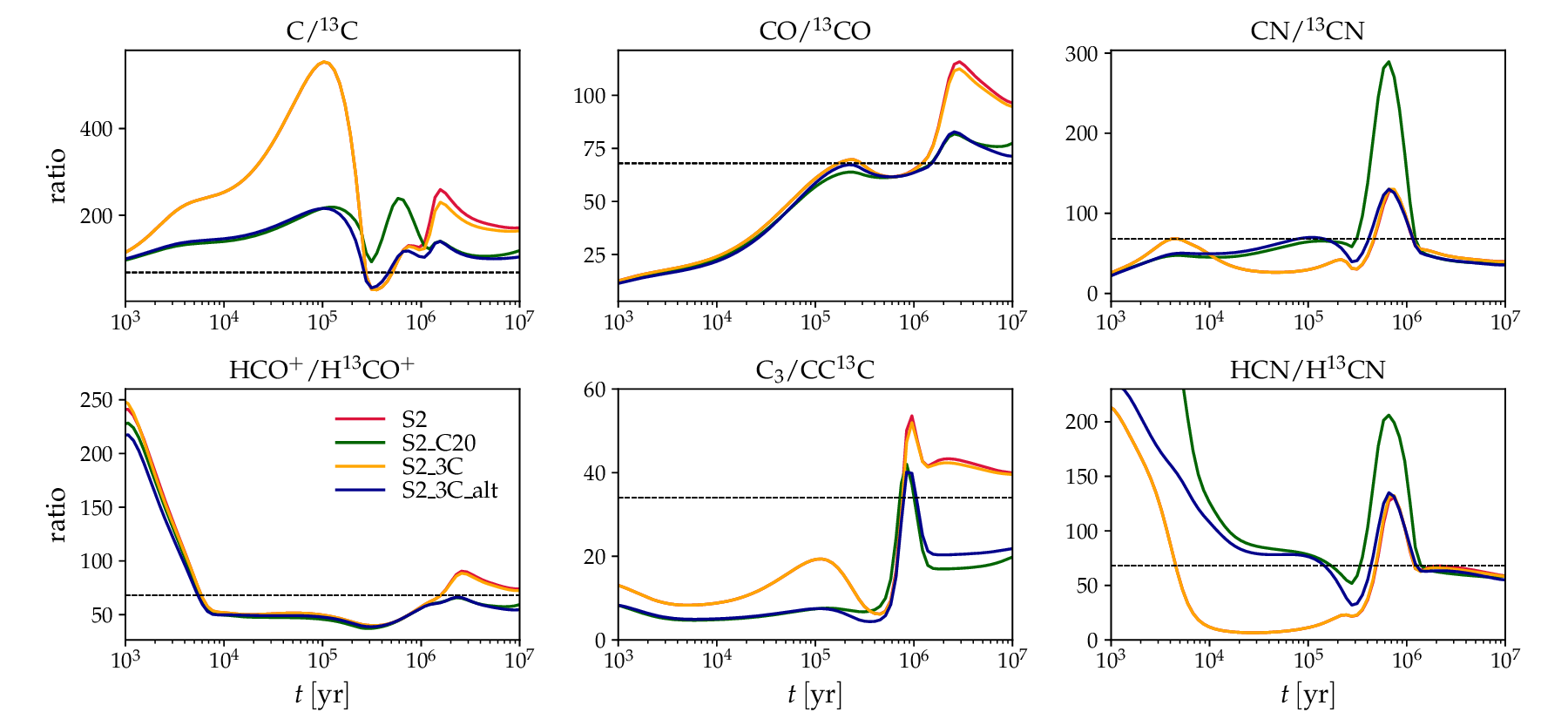}
    \caption{Selected $\cc/\cz$ abundance ratios as a function of time in the variants of simulation~S2 (S2, red; S2\_C20, green; S2\_3C, orange; S2\_3C\_alt, blue). The dashed black line in each panel indicates the elemental $\rm ^{12}C$/$\rm ^{13}C$ ratio (68), except for $\rm C_3$ for which a ratio of 34 is used as our model does not distinguish between $\rm CC\cz$ and $\cz CC$. The results of simulations~S2~and~S2\_3C overlap almost perfectly.}
    \label{fig:CfracComparison}
\end{figure*}

\begin{figure*}
\centering
        \includegraphics[width=2.0\columnwidth]{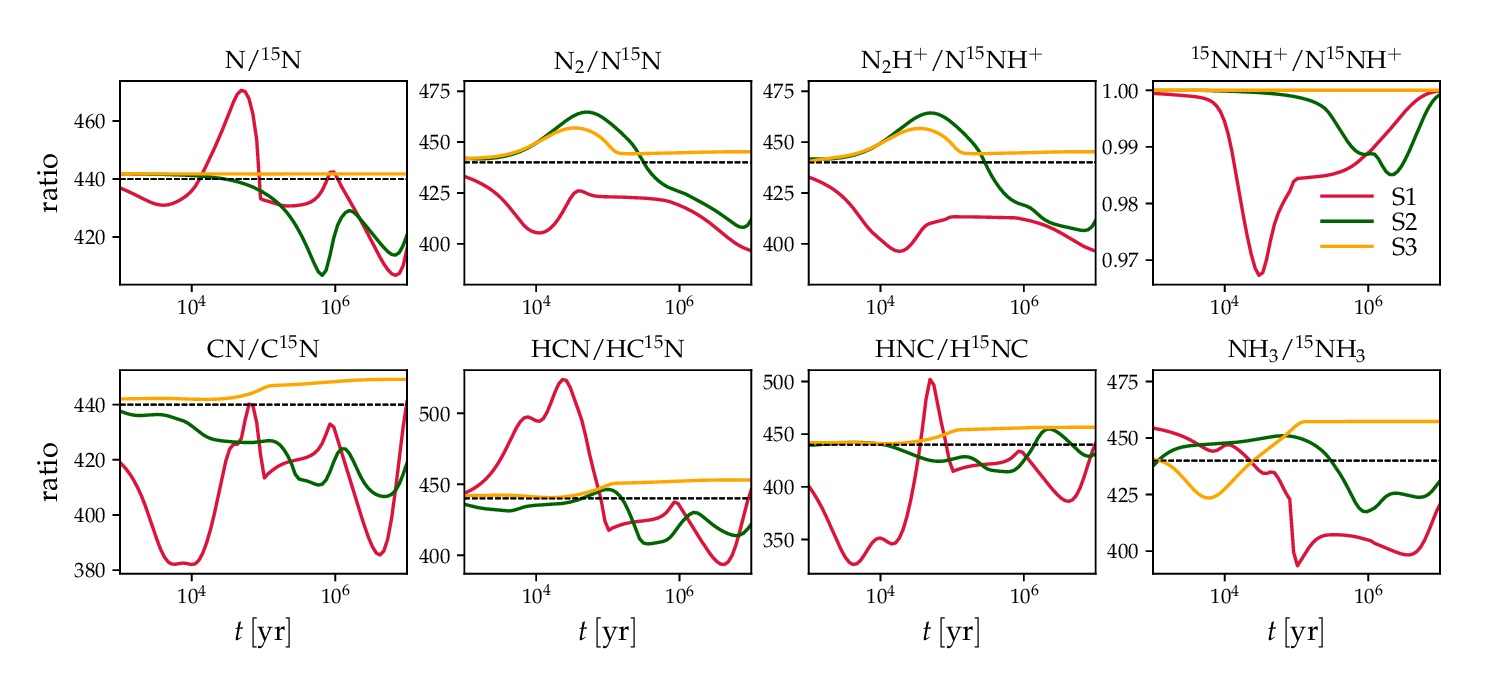}
    \caption{Selected $\nn/\nz$ abundance ratios as a function of time in simulations S1~to~S3, labeled in the top right panel. The dashed black line in each panel indicates the assumed elemental $\rm ^{14}N$/$\rm ^{15}N$ ratio (440).}
    \label{fig:Nfrac}
\end{figure*}

As it was already pointed out in Sect.\,\ref{ss:chemicalNetwork}, the present isotopic chemical networks contain several differences to those used in C20, also for carbon chemistry. Therefore, it is important to check whether the current model yields results similar to those of C20, or whether the results deviate from one another due to the modifications made for this paper. As the models differ in several ways, we analyze the effect of the main features of the present model in a sequential fashion, using modified versions of the S2 simulation (see Table~\ref{tab:simulationParameters}).

Figure~\ref{fig:CfracComparison} shows a comparison of the simulation results as predicted by the S2 simulation variants, for several species also plotted by C20. Let us first examine the isotopic abundance ratios in the C20 model\footnote{The simulations using the C20 network have been rerun for this paper, that is, the curves shown in Fig.\,\ref{fig:CfracComparison} are not a reproduction of the corresponding results shown in C20. Even though the chemical network is the same, some differences to the results of C20 can be seen. This is because of $\rm H_2$ self-shielding which was omitted in C20 but is included here (following \citealt{Draine96}, Eq.\,37).} (simulation S2\_C20; green) and the present model with only carbon fractionation and using the same exchange reactions as in C20 (simulation S2\_3C\_alt; blue). We reiterate that the difference between these models is that the latter tracks the position of the $\cz$ atoms in the various molecules, while the former does not. As expected, this modification does lead to differences in predicted abundance ratios, which can be pronounced in some cases (for example, C, CN). However, even the largest differences between the results of the two simulations are at a factor of a few level, and the present expanded model is not in conflict with the main conclusions of C20.

Introducing new exchange reactions for carbon fractionation into the network (simulation S2\_3C; orange) has a large impact on the fractionation of atomic carbon especially at early times in the simulation, though also at late times the $\cc/\cz$ ratio is boosted by a factor of $\sim$two compared to C20. This boost is due to the newly added exchange reactions involving $\rm C_3$. Simultaneously, the $\rm HCN/H\cz N$ ratio obtains a low value ($\sim$10) from $t = 10^4$ to a few $\times 10^5 \, \rm yr$ (similarly for HNC, not shown in the Figure), which is mainly because of efficient conversion of $\rm HCN$ to $\rm H\cz N$ via exchange reaction~(24) (see Table~\ref{tab:reaccolzi}). The fact that this ratio lies relatively far from 68 has observational implications, which we discuss briefly in Sect.\,\ref{ss:C+N}. The results of simulation~S2 are in agreement with the gas-grain model of \citet{loison2020} to within a factor of $\sim$two (the overall agreement is worse in the simulations S2\_C20 and S2\_3C\_alt), demonstrating that using an extensive set of exchange reactions is required to improve the accuracy of the simulation of isotopic fractionation (though naturally within the accuracy of the reaction rates themselves). 

Our results confirm the statements of \citet{loison2020} regarding the important role of $\rm C_3$ for carbon fractionation. Further theoretical and experimental work needs to be carried out to constrain the rate coefficients of exchange reactions involving $\rm C_3$. \citet{loison2020} reiterated the particular importance of the $\rm C_3 + O$ reaction (see also the references in \citealt{loison2020}), and tested the effect of order-of-magnitude variations in the rate coefficient of this reaction on the $\cc/\cz$ ratios of various species. They pointed out that the rate coefficient is poorly constrained and that its value depends strongly on how it is computed. To our knowledge, no subsequent progress on constraining this reaction has been made, and therefore we have not explored this issue further in this work.

Finally, adding nitrogen fractionation and time-dependent spin-state chemistry to the reaction scheme, that is, using the full model developed for this work (simulation S2; red), has little influence on carbon fractionation. This means that the newly included exchange reactions combining carbon and nitrogen fractionation do not play a large role in the chemistry of $\cz$, and secondly this finding confirms the expectation that spin-state chemistry does not significantly influence carbon chemistry either.

\subsection{Nitrogen isotopologues}\label{ss:nitrogenIsotopologues}

Figure~\ref{fig:Nfrac} shows selected isotopic abundance ratios as a function of time obtained from simulations S1 to S3. There is clear variation in the fractionation level between the various N-bearing species, also depending on the physical conditions.

In simulation S1, which mimics the physical conditions in the center of a prestellar core, the $\rm N_2H^+$/$\rm N\nz H^+$ ratio is almost constant as a function of time, remaining near the elemental $\nn/\nz$ ratio of 440. There is also little variation between the two singly substituted forms of $\rm N_2H^+$, with the $\rm ^{15}NNH^+$/$\rm N\nz H^+$ ratio remaining near unity at all times. As expected, the $\rm N_2H^+$ $\nn/\nz$ ratio is very close to that of $\rm N_2$, while the atomic $\nn/\nz$ ratio presents minor fluctuations around the elemental ratio. CN and its derivatives, HCN and HNC, present similar behavior at late simulation times, showing enrichment in $\nz$. There is an early-time difference between the $\rm HCN/HC\nz$ and $\rm HNC/H\nz C$ ratios, however, which is due to a combination of several reactions that leads to a difference in the early-time $\rm HCN/HNC$ and $\rm HC\nz/HC\nz$ ratios. The ammonia $\nn/\nz$ ratio also stays quite close to the elemental value, with slight anti-fractionation at late simulation times. Our results are broadly similar to those presented by \citet{Wirstrom18} who adopted physical conditions close to simulation S1, despite the fact that there are many differences between the model setups, for example that we are presently considering grain-surface chemistry and desorption processes. We confirm the finding of \citet{Wirstrom18} that the observed nitrogen fractionation ratios of $\rm N_2H^+$ cannot be explained by means of updated rate coefficients for the exchange reactions included in current models.

\begin{figure*}
\centering
        \includegraphics[width=2.0\columnwidth]{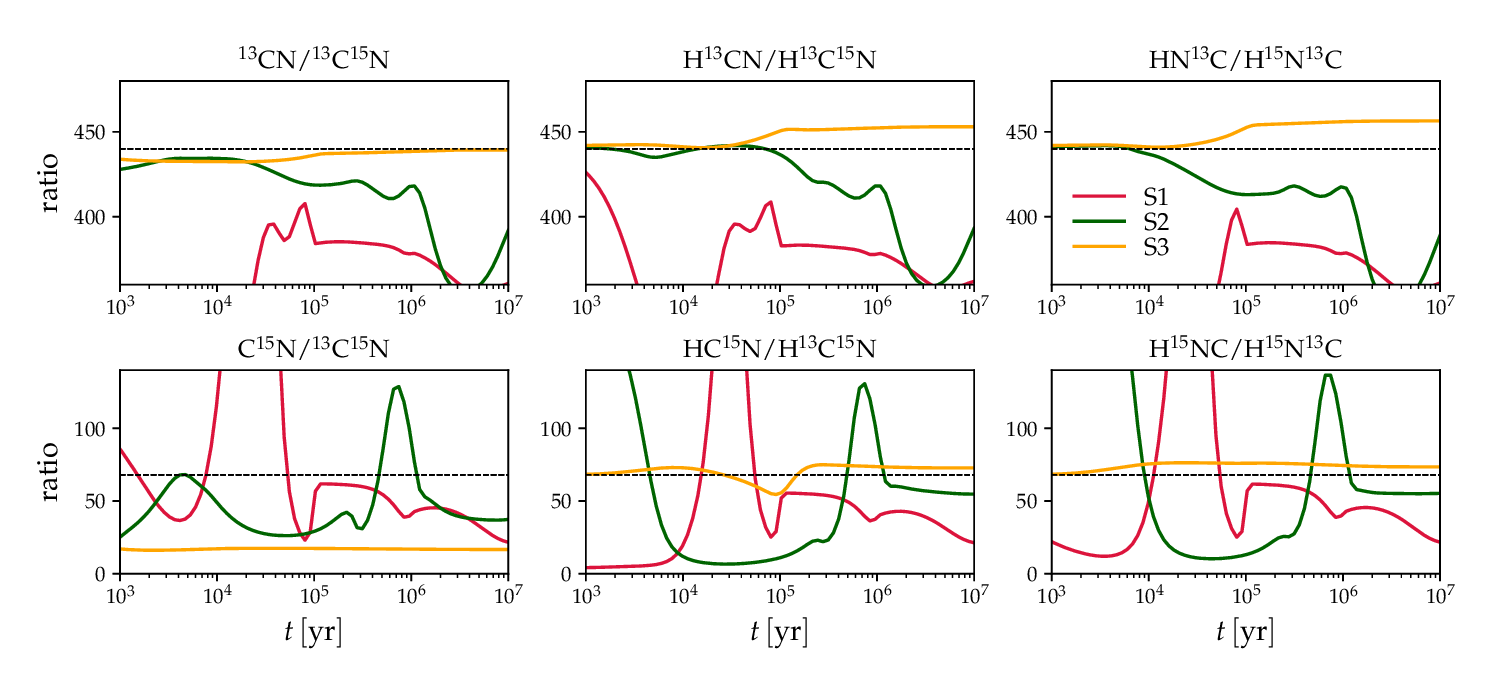}
    \caption{Selected abundance ratios as a function of time in simulations S1~to~S3, labeled in the top right panel. {\sl Top:} ratios of $\rm ^{13}C$-substituted CN, HCN, and HNC to their $\cz$+$\nz$-substituted forms.  The dashed black line in each panel indicates the assumed elemental $\nn/\nz$ ratio (440). {\sl Bottom:} ratios of $\nz$-substituted CN, HCN, and HNC to their $\cz$+$\nz$-substituted forms. The dashed black line in each panel indicates the assumed elemental $\cc/\cz$ ratio (68).}
    \label{fig:N+Cfrac}
\end{figure*}

There are some differences in the various ratios in simulation S2 compared to S1, which is to be expected given the difference in the physical conditions. However, the results from the two simulations are overall similar -- none of the molecules shows a significant deviation from the elemental $\nn/\nz$ ratio at late simulation times. The physical conditions in simulation~S2 are very similar to those considered by \citet{loison2019}, and although we use different initial conditions for the chemistry and -- importantly for nitrogen chemistry -- consider the time-dependent variation in the $\rm H_2$ OPR, we recover very similar abundance ratios to those presented by \citet{loison2019} (see Sect.\,\ref{ss:H2opr} for more discussion on the effect of the $\rm H_2$ OPR on our results).

In simulation~S3, which corresponds to a lower density and higher temperature, and much lower visual extinction compared to simulations S1~and~S2, there are generally only weak deviations from the elemental $\nn/\nz$ ratio. This is to be expected because the higher gas temperature helps most exchange reactions to proceed efficiently in each direction, preventing a large degree of fractionation. Indeed, the overall trend is that the various $\nn/\nz$ ratios present a slight enhancement over the elemental ratio.

In summary, our model predicts generally small deviations, on the order of ten percent, for the various $\nn/\nz$ ratios from the elemental ratio. Notably, for any of the physical conditions considered here, the model does not reproduce the extremely high anti-fractionation of $\rm N_2H^+$ observed by \citet{Bizzocchi13} in the prestellar core L1544 ($\nn/\nz \sim 1000$) and upward of $\sim 700$ in a sample of prestellar cores observed by \citet{Redaelli18}. Possible explanations for this discrepancy between the models and the observations are the trapping of $\nz$, and in particular of $\nz N$, onto grain surfaces already before the formation of the prestellar core \citep{Furuya18}, or the suggested large isotope-dependent difference in $\rm N_2H^+$ dissociative recombination rate \citep{loison2019,Hily-Blant20}.

\subsection{Species containing both $\cz$ and $\nz$}\label{ss:C+N}

Figure~\ref{fig:N+Cfrac} presents the $\rm \cz N/\cz\nz$ and $\rm C\nz/\cz\nz$ abundance ratios for CN, HCN, and HNC in the three fiducial simulations. For the $\rm C\nz/\cz\nz$ ratios, there are no obvious trends and all simulations predict varying degrees of deviation from the statistical ratio of 68. The results are highly time-dependent and there are transient variations in many ratios. These transient variations are the largest in simulation~S2 which is in line with the results of C20, who used the same physical conditions as in our simulation~S2, and found similarly large transient effects. For the $\rm \cz N/\cz\nz$ ratios, however, the simulations predict increasing fractionation with volume density, with the ratios decreasing below 400 in simulation~S1. This behavior is very different to the singly fractionated nitrogen ratios (Fig.\,\ref{fig:Nfrac}), for which there are no clear trends as a function of the physical conditions. The $\cz$+$\nz$ fractionation ratios are sensitive to the newly introduced exchange reactions; we examine their effect in more detail in Sect.\,\ref{ss:doubleSubstitution} below.

\begin{figure}
\centering
        \includegraphics[width=0.9\columnwidth]{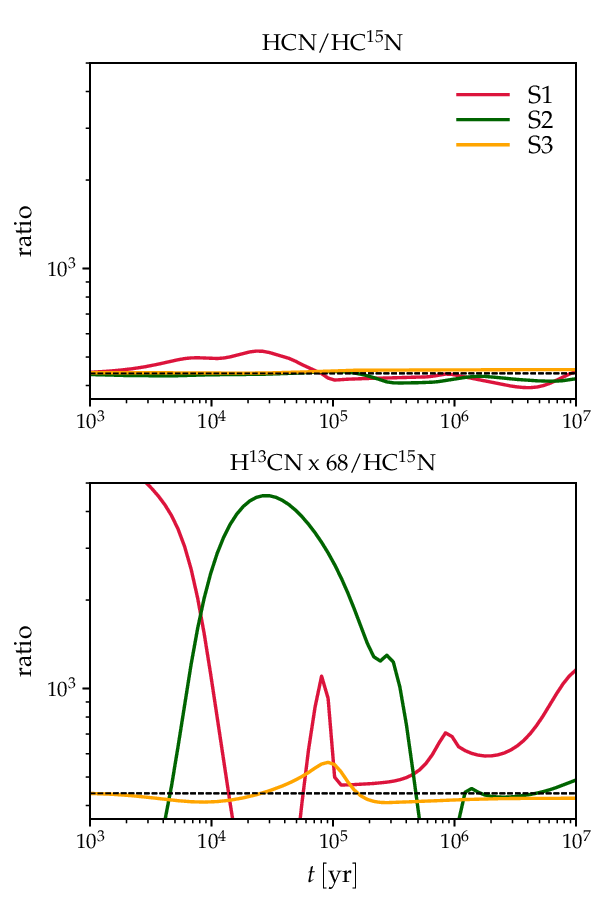}
    \caption{Time evolution of the $\rm HCN/HC\nz$ ratio in simulations~S1~to~S3, labeled in the top panel (data reproduced from Fig.\,\ref{fig:Nfrac}). The top panel displays the ratio as predicted directly in the simulations, while in the bottom panel, the HCN abundance has been derived by scaling the simulated $\rm H\cz N$ abundance by a factor of 68. The logarithmic scaling of the y-axis in both panels was employed here to accentuate the magnitude of the differences between the results. The dashed black line in each panel indicates the assumed elemental $\nn/\nz$ ratio (440).}
    \label{fig:HCN_scaling}
\end{figure}

Although we will explore the observational implications of our new isotope chemical model in more detail in a future dedicated study, we briefly demonstrate in Fig.\,\ref{fig:HCN_scaling} the importance of tracking the isotopic ratios time-dependently, as opposed to converting observationally deduced abundances using a constant scaling factor. The figure shows the evolution of the $\rm HCN/HC\nz$ ratio as predicted directly by the present model and as obtained by scaling the model prediction for $\rm H\cz N$ by a constant factor of 68. Evidently, there can be a huge discrepancy in the $\rm HCN/HC\nz$ ratio resulting from the two approaches, especially at higher volume densities -- the scaled ratio is in model~S2 up to an order of magnitude different to the elemental $\nn/\nz$ ratio depending on the evolutionary time. The peak is caused by the $\rm HCN/H \cz N$ ratio being less than 10 in the time interval from $\sim$$10^4$\,yr to a few $10^5 \, \rm yr$ (cf. Fig.\,\ref{fig:CfracComparison}), hence giving rise to a high HCN abundance when the $\rm H \cz N$ abundance is scaled by a constant factor of 68. The implications of this finding are complicated by the fact that observations typically trace only a part of the total distribution of a molecule, and optical depth effects may also mask the part of the distribution that is prone to the large fluctuations. Nevertheless, it is clear that dedicated simulations are required to support the interpretation of observations, and that assuming constant elemental abundance ratios even within a molecular cloud may lead to large errors.

\section{Discussion}\label{s:discussion}

\subsection{Effect of new $\cz$ + $\nz$ exchange reactions}\label{ss:doubleSubstitution}

\begin{figure*}
\centering
        \includegraphics[width=2.0\columnwidth]{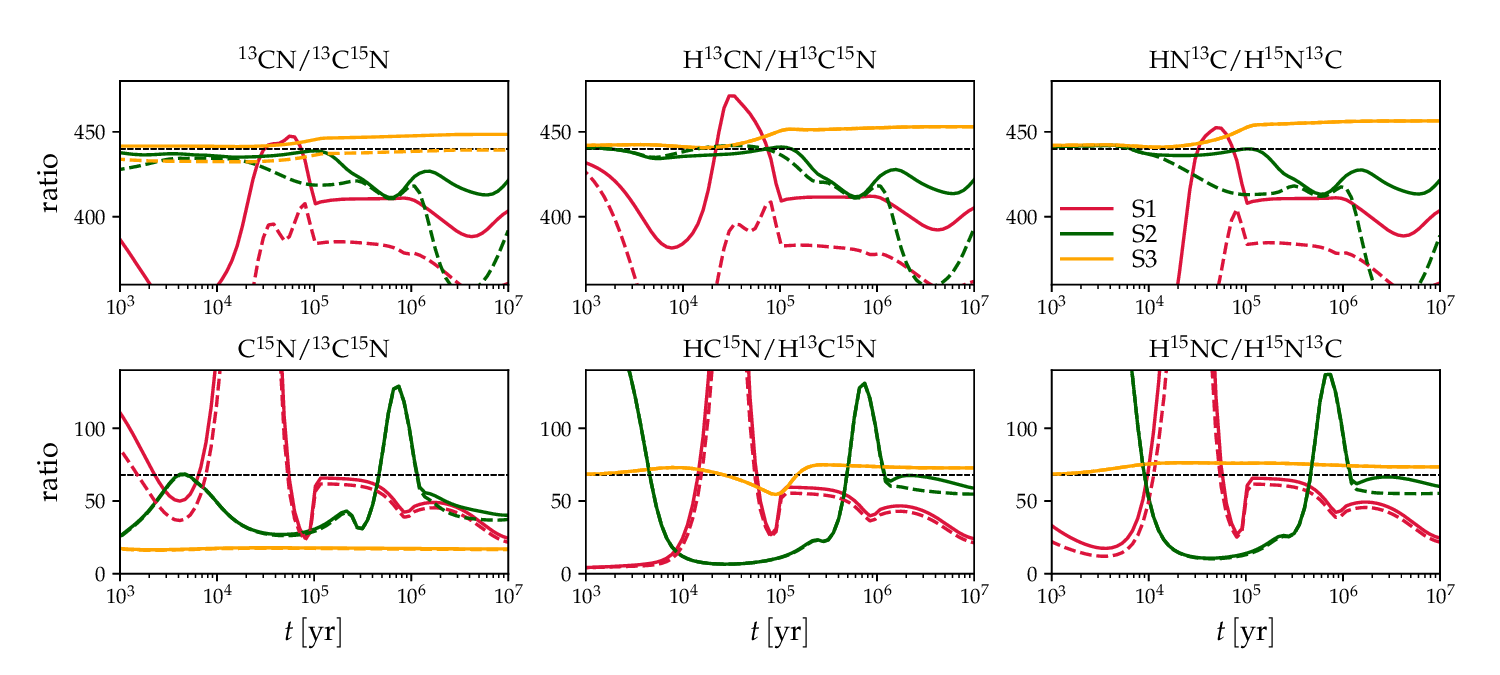}
    \caption{Selected abundance ratios (indicated on top of each panel) as a function of time in simulations S1~to~S3, labeled in the top right panel. Solid lines correspond to simulations where the N+C exchange reactions presented in Table~\ref{tab:reaccolzi2} are excluded from the chemical network, while dashed lines show the corresponding results in the fiducial simulations that include all exchange reactions (reproducing data from Fig.\,\ref{fig:N+Cfrac}). The dashed black line in each panel indicates either the assumed elemental $\cc/\cz$ or $\nn/\nz$ ratio, depending on the quantity being plotted.}
    \label{fig:N+Cfrac_noNewExchange}
\end{figure*}

We have introduced in this work a set of exchange reactions involving molecules fractionated in both $\cz$ and $\nz$ (Table~\ref{tab:reaccolzi2}) that have not been considered in previous models of isotopic fractionation. Associated with common molecules and presenting barriers of several tens of Kelvin in many cases, they may also have an influence on the chemical evolution of singly fractionated molecules besides their influence on molecules fractionated in both C and N. We assess the importance of their inclusion in Fig.\,\ref{fig:N+Cfrac_noNewExchange}, which shows the same abundance ratios as in Fig.\,\ref{fig:N+Cfrac} but using either all of the exchange reactions given in Table~\ref{tab:reaccolzi2}, or excluding the new $\cz$+$\nz$ exchange reactions. The ratios, involving doubly fractionated molecules, show dependence on the included exchange reactions -- without the reactions, there is less fractionation (which is to be expected) and the various $\rm \cz N/\cz\nz$ ratios tend to be closer to the assumed elemental ratio of 440. The differences are only prominent in the highest-density simulation (S1), however, and are even there only on the 10\% level. The exchange reactions involving doubly fractionated species have only a very minor influence on single fractionation (the changes are slight for $\nz$-bearing species, and virtually non-existent for $\cz$-bearing species), and these are not shown here in the main text for the sake of brevity -- additional simulation results are shown in Appendix~\ref{appendix:a}.

\subsection{Isotope-selective self-shielding of $\rm N_2$ and CO}\label{ss:N2selfShielding}

\begin{figure*}
\centering
        \includegraphics[width=1.8\columnwidth]{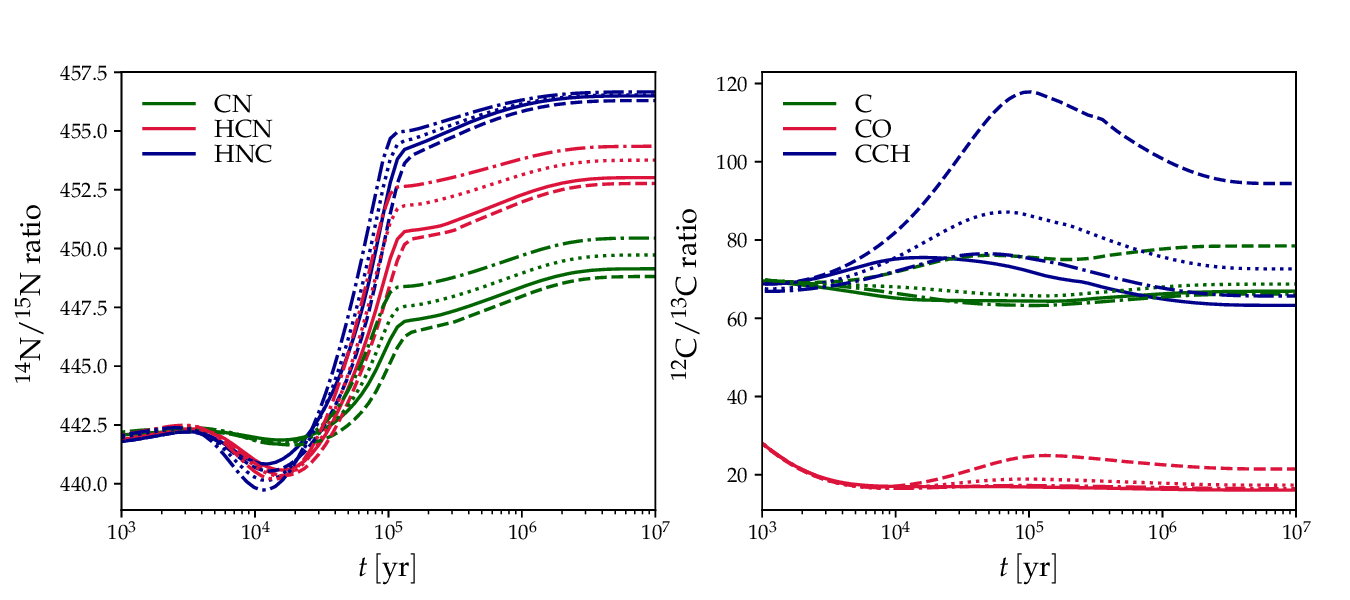}
    \caption{$\nn/\nz$ ratios of CN, HCN, and HNC as a function of time in three simulations: S3 (solid lines), S3\_ss (dashed lines), S3\_ss with $G_0 = 2$ (dotted lines), and S3\_ss with $G_0 = 3$ (dash-dotted lines).}
    \label{fig:NC_selfShielding}
\end{figure*}

Recently, \cite{Spezzano22b} presented observations of the $\nn/\nz$ ratios of CN, HCN, and HNC toward the prestellar core L1544 and found $\nn/\nz$ gradients across the core. For example for HNC, they find that the $\nn/\nz$ ratio increases toward the southeast where the core is more exposed to the interstellar radiation field (ISRF) \citep{Spezzano16b, Spezzano22b}. However, CN and HCN present a reverse trend in their data set. One possible explanation for the existence of the gradients is self-shielding, which is expected to be weaker for $\rm N^{15}N$ than for $\rm N_2$ in the outer core regions due to the much lower column density of the former. This would imply enhanced photo-destruction of $\rm N\nz$, releasing atomic $\nz$ that can be subsequently incorporated into other molecules and hence influencing the $\nn/\nz$ ratios \citep{Furuya18, Visser18}.

To test the effect of self-shielding on our results on a qualitative level, we ran a variant of the S3~simulation, S3\_ss, where the self-shielding of $\rm N_2$, $\rm N\nz$, CO, and $\rm \cz O$ is taken into account (this is not done in any of the other simulations presented in Table~\ref{tab:simulationParameters}). For this, we estimated the appropriate $\rm N_2$ and CO column densities by first taking the corresponding abundances from the fiducial S3 simulation at a late evolutionary time ($\sim$$10^6\,\rm yr$). These were multiplied by the thickness of the virtual envelope\footnote{The simulations are zero-dimensional, and hence there is no actual attenuating gas included in the simulation.} as determined from the external visual extinction (1~mag in simulation~S3) using the relation $N({\rm H_2})/A_{\rm V} = 10^{21} \, \rm cm^{-2}\,mag^{-1}$, finally yielding $N({\rm N_2}) = 2\times10^{13} \, \rm cm^{-2}$ and $N({\rm CO}) = 9\times10^{14} \, \rm cm^{-2}$. The self-shielding factors for $\rm N_2$ and $\rm N\nz$ photodissociation were taken from \citet{Li13} and \citet{Heays14}, while those for CO and $\rm \cz O$ were taken from \citet{Visser09}. The data are given as a function of $\rm N_2$ (CO) column density, and so the $\rm N\nz$ ($\rm \cz O$) column density does not need to be separately estimated (the literature data assumes $N({\rm N_2})/N({\rm N}\nz) = 225$ and $N({\rm CO})/N({\rm \cz O}) = 69$).

Figure~\ref{fig:NC_selfShielding} displays the time evolution of the CN, HCN, and HNC $\nn/\nz$ ratios in the S3 and S3\_ss simulations. The difference between the simulation results is almost negligible, with only less than a 1\% difference between the abundance ratios predicted by the two simulations. Motivated by the observational evidence for nonuniform illumination of L1544 by the ISRF, we ran two additional simulations like S3\_ss but where the strength of the ISRF is increased from the fiducial value ($G_0 = 1$) by a factor of two or three (but keeping the same assumed $\rm N_2$ column density to compute the self-shielding). The results of this simulation are also shown in Fig.\,\ref{fig:NC_selfShielding}. We find that all three $\nn/\nz$ ratios increase when the external illumination grows in strength. HNC does present a slightly different response compared to CN and HCN in that its $\nn/\nz$ ratio increases less when $G_0$ is increased, but the overall trend is obviously different compared to what \citet{Spezzano22b} observed in L1544. Also, even with the increased amount of external illumination, and hence a larger emphasis on the difference between self-shielding of $\rm N_2$ and $\rm N\nz$, the simulation results still show only one percent level differences to the fiducial S3 simulation.

Figure~\ref{fig:NC_selfShielding} also shows the $\cc/\cz$ ratios of atomic C, CO, and CCH. Here the differences between the simulation results are larger than in the case of nitrogen, but still within a factor of two of the fiducial model results. First, we see that when self-shielding is introduced, all of the ratios increase in magnitude. This arises because in the physical conditions of model~S3 (and hence S3\_ss), the main $\cc$ and $\cz$ carriers are $\cc^+$ and $\cz^+$, which are created (at late simulation times) by photodissociation of CO ($\rm \cz O$) followed by photoionization of atomic $\cc$ ($\cz$). Reactions involving $\rm C^+$ (cf. Table~\ref{tab:reaccolzi}, especially Reaction~(1)) are then the most important ones determining the isotopic fractionation. When self-shielding is introduced, the decreased production of ionized carbon, in particular of $\cz^+$, decreases the overall level of fractionation and the $\cc/\cz$ ratios rise with respect to the fiducial model~S3.

However, the effect of the strength of the ISRF on the results is not straightforward, and the various $\cc/\cz$ ratios actually decrease when the strength of the ISRF is increased, which seems counterintuitive at first. The differential photodissociation efficiency of CO and $\rm \cz O$ due to self-shielding (CO self-shields more efficiently than $\rm \cz O$ owing to its larger column density) and hence the differential production of $\cc^+$ and $\cz^+$ has a more pronounced effect in stronger radiation fields -- when $G_0$ is increased, the abundance of $\cz^+$ increases more than that of $\cc^+$. This leads to an increase in fractionation, that is, a decrease in the various $\cc/\cz$ ratios. We stress that the effect is indeed due to self-shielding. A comparison of the results of model~S3 run with different $G_0$ factors (not shown) does not reveal significant variations in the $\cc/\cz$ ratios.

The self-shielding coefficients for $\rm N_2$ and $\rm N\nz$ are, for our chosen values of $\rm H_2$ and $\rm N_2$ column density, very similar to each other, with only a $\sim$20 percent difference between them. This translates to a marginal difference in the $\rm N_2$ and $\rm N\nz$ photodissociation rates, and hence the effect on the atomic $\rm N/\nz$ ratio is in our tests much smaller than what has been observed by \citet{Spezzano22b}. It is conceivable that for different combinations of $\rm H_2$ and $\rm N_2$ column density, we might find different trends in the CN, HCN, and HNC $\nn/\nz$ ratios. Similar arguments apply to the analysis of the $\cc/\cz$ ratios. More detailed modeling is required to quantify this and to reach robust conclusions as regards the comparison to the observations. Also, we note that the observed spatial variations in the $\nn/\nz$ ratios may be due not only to selective self-shielding, but also to local variations in the $\cc/\cz$ ratio in the molecules. Based on the present simulation results, the latter effect might be dominant. (\citealt{Spezzano22b} assumed a constant $\cc/\cz$ ratio of 68 for all molecules in their sample.) We will investigate this topic in more detail in a future work, which will include detailed prestellar core simulations that allow for a quantitative assessment of the effects of localized variations in the $\cc/\cz$ ratio. In that work, we will also investigate the abundance gradients of other molecules that are implied by the results of simulations S1~to~S3.

\subsection{Effect of $\rm H_2$ OPR}\label{ss:H2opr}

\begin{figure*}
\centering
\begin{picture}(500,200)(0,0)
\put(-15,0){
\begin{picture}(0,0) 
        \includegraphics[width=1.0\columnwidth]{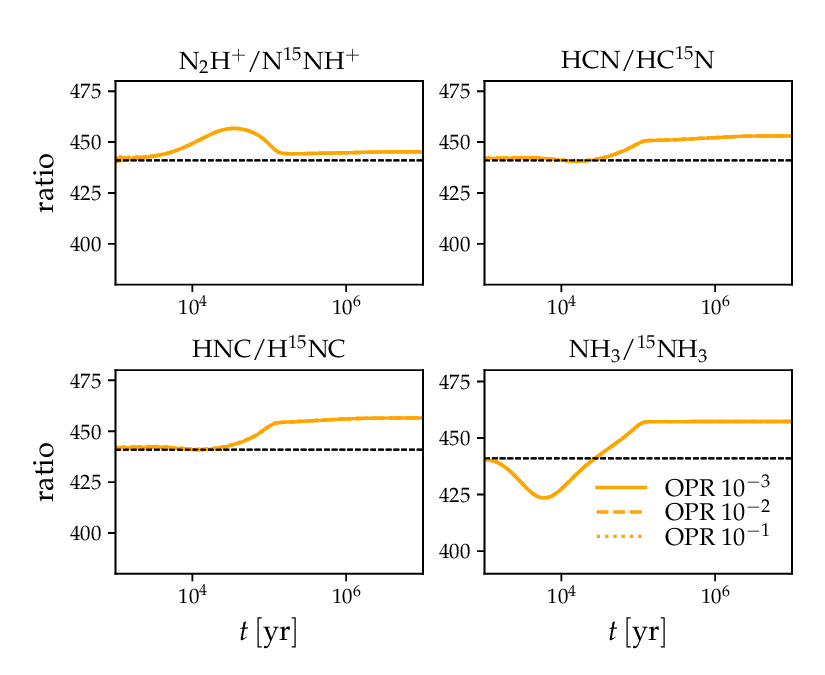}
\end{picture}}
\put(270,15){
\begin{picture}(0,0) 
        \includegraphics[width=0.8\columnwidth]{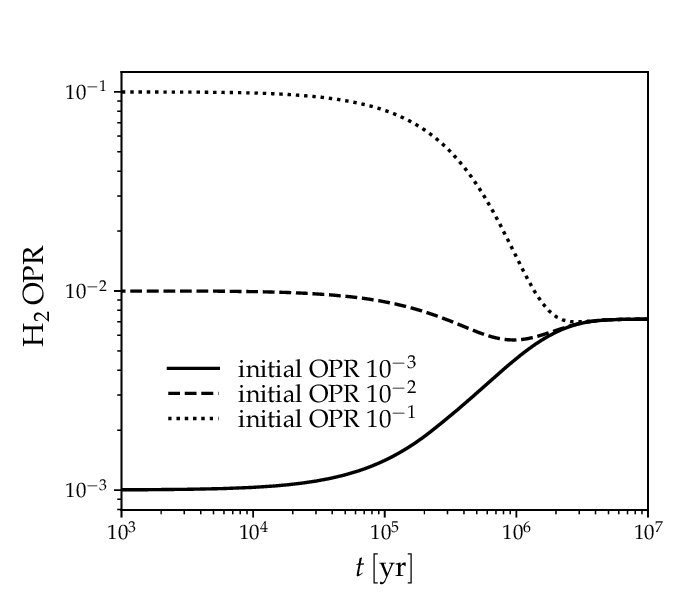}
\end{picture}}
\end{picture}  
    \caption{Effect of the $\rm H_2$ OPR on the simulation results. {\sl Left:} selected $\nn/\nz$ abundance ratios as a function of time in simulation~S3 for three different initial $\rm H_2$ OPR values, labeled in the plot. The dashed black line in each panel indicates the elemental $\rm ^{14}N$/$\rm ^{15}N$ ratio (440). {\sl Right:} time evolution of the $\rm H_2$ OPR in simulation~S3.}
    \label{fig:opTests}
\end{figure*}

The simulation results presented in the preceding sections all assume a low initial $\rm H_2$ OPR of $10^{-3}$. Higher initial values can however be expected in warmer, low-density media such as in the S3 simulation considered here. We have thus tested two additional values of the initial OPR ($10^{-2}$ and $10^{-1}$) on the results of simulation~S3. Even though $\rm H_2$ is expected to form on grains following the high-temperature statistical OPR of 3 \citep{Watanabe10}, the gas-phase OPR deviates from the thermal ratio unless the volume density is very low \citep{Lupi21} due to the competition between gas-phase exchange reactions and surface formation, and hence an initial gas-phase ratio of above $\sim$$10^{-1}$ is not appropriate for the present simulations. Because carbon chemistry is not expected to be dependent on the $\rm H_2$ OPR, we focused in these tests on nitrogen-bearing species only.

Figure~\ref{fig:opTests} shows the time evolution of selected $\nn/\nz$ abundance ratios assuming the three different initial $\rm H_2$ OPR values in the S3~simulation. Evidently, the initial value of the OPR has virtually no influence on the results at late simulation times -- differences in the ammonia abundances for example do arise, but only from the beginning of the simulation up to times on the order of 100\,yr (not shown in the Figure). This means that the initial OPR value plays only a minor role for the abundances of $\rm NH_3$ and $\rm \nz H_3$. The value of the rate coefficient of the $\rm N^+ + H_2 \rightarrow NH^+ + H$ reaction is the determining factor for the ammonia abundance, regardless of the inclusion or exclusion of the spin states in the simulation. We have confirmed this by running an alternative S3~simulation with a chemical network that is otherwise identical to our fiducial one but where the spin states have been completely suppressed. For this test we used the rate coefficient of reaction~(41) in Table~\ref{tab:reaccolzi2} to represent the rate coefficient of $\rm N^+ + H_2 \rightarrow NH^+ + H$ (see also \citealt{Dislaire12}). This is equivalent to assuming that all $\rm H_2$ is in ortho form. Indeed, the results of the test simulation are nearly identical to simulation~S3 with $\rm OPR = 10^{-1}$, also for the early time evolution.

We have also run similar $\rm H_2$ OPR tests applied to simulations~S1~and~S2, and in those simulations we do see differences for example in the late-time $\rm NH_3/\rm \nz H_3$ ratio depending on the initial OPR, but the differences are only on the $\sim$10 percent level. This agrees with the results of \citet{Roueff15}, who found only small deviations in the nitrogen fractionation ratios when the OPR was varied, using a gas-phase model with an OPR value fixed in time. In conclusion, we expect only a very weak dependence of the isotopic abundance ratios of nitrogen-bearing molecules on the initial $\rm H_2$ OPR even when the OPR is allowed to evolve with time.

\subsection{Uncertainty in the rate coefficients of fractionation reactions involving $\rm C_2$}

\citet{Rocha21} have recently presented new rate coefficients for reactions involving $\cz$ and $\rm C_2 / C\cz$. Their rate coefficients are decidedly lower than what we employ for the corresponding reactions (Table~\ref{tab:reaccolzi}) despite the nearly identical exothermicity values (their Table~3). For example for our reaction~(5) in the forward direction ($^{13}$C + C$_{2}$ $\rightarrow$ $^{12}$C + $^{13}$CC), we obtain a rate coefficient of $2.89\times 10^{-10} \, \rm cm^{-3}$ at 10\,K, whereas the fit of \citet{Rocha21} gives $1.52\times 10^{-11} \, \rm cm^{-3}$ --  though it is to be noted that the rate coefficients of \citet{Rocha21} were calculated for higher temperatures (upward of 25\,K) than we consider here (10\,K in most simulations).

We tested the effect of switching to the \citet{Rocha21} rate coefficients for their reactions~(2)~and~(3) (with the $\rm C_2$ variants in the $X^1\Sigma_g^+$ state) by running new variants of simulations S1, S2, and S3 with the rate coefficients of the reactions in question switched to the \citet{Rocha21} values. We found that the effects on the simulation results are highly dependent on the physical conditions. In model~S3, the rate coefficients of these reactions play only a very minor role, but in the high-density and low-temperature environment of model~S1 we obtain variations of up to an order of magnitude in the $\cc/\cz$ ratios of $\rm C_2$ (and typically a factor of a few for other ratios) depending on the choice of the rate coefficients. The reason why the \citet{Rocha21} coefficients are so much lower than ours, or those of \citet{loison2020}, is unknown at present and hence we have not included their rate coefficients in the bulk of the analysis presented here. However, these tests highlight that constraining the rate coefficients of critical reactions is of great importance for the predictions of chemical models, and further experimental and theoretical work on this is called for.

\section{Conclusions}\label{s:conclusions}

We presented a new gas-grain model for the combined fractionation of $\cz$ and $\nz$ in the ISM\footnote{The chemical networks developed for this study can be obtained upon request to the corresponding author.}. Our model is the first to include a time-dependent treatment for spin-state chemistry in the context of $\cz$ and $\nz$ fractionation, which is important for simulating the chemistry of nitrogen. In addition, we introduced a set of new exchange reactions for $\cz$+$\nz$ fractionation, and provided updates to the rate coefficients of several reactions involving $\cz$. We applied the new model to a set of physical conditions representing the environment across a prototypical prestellar core, from its strongly shielded dense and cold center to the warmer outer regions in the envelope where the interstellar radiation field still penetrates.

We found that the $\cc/\cz$ ratio in the various molecules can deviate by a factor of several from the elemental $\cc/\cz$ ratio, depending on the physical conditions and on the time. For nitrogen, however, we obtained much smaller deviations from the elemental ratio, on the order of ten percent. We also studied the double fractionation ratios ($\rm \cz N/\cz\nz$ and $\rm C\nz/\cz\nz$), finding also in these cases that the $\cc/\cz$ ratios are highly variable. In contrast to the single fractionation case, we found a trend in the $\rm \cz N/\cz\nz$ ratios such that the ratios decrease with increasing volume density, though still remaining within a few tens of percent from the elemental $\nn/\nz$ ratio. We also tested the effect of selective self-shielding of $\rm N_2$ and CO versus $\rm N\nz$ and $\rm \cz O$, respectively, on isotopic fractionation at low volume densities. For nitrogen, self-shielding appears to play only a minor role for the isotopic abundance ratios of HCN and HNC, for example. For carbon, larger deviations from the fiducial models are seen when the self-shielding is included in the simulation, though the $\cc/\cz$ ratios remained within a factor of two from their fiducial values. It must be emphasized, however, that our tests were only qualitative and very limited owing to the zero-dimensional nature of the simulations.

The isotopic abundance ratios predicted by our new model are generally in a factor of two or better agreement with those of similar models in the literature. For example, we recover the large molecule-dependent deviations from the elemental $\cc/\cz$ ratio predicted by \citet{Colzi20}, although we obtained a substantially larger degree of carbon fractionation at low volume densities owing to a lower amount of external extinction assumed in the present model. Also, for canonical dark cloud conditions, we obtained similar single-fractionation ratios to the recent works by \citet{loison2019} and \citet{loison2020}, and our results support the finding of \citet{Wirstrom18} that updates to the rate coefficients of the presently considered isotopic exchange reactions are not enough to reproduce the observed (anti-)fractionation of several N-bearing species.

The simulation results are sensitive to the physical conditions, and imply gradients in isotopic fractionation across prestellar cores. Such gradients were recently deduced observationally by \citet{Spezzano22b}, but the interpretation of the observations is limited by the nonavailability of quantitative information on the gradients from a simulation point of view. The analysis is further complicated by the fact that the various isotopic forms do not necessarily trace the same regions within the core. We will carry out more detailed modeling to quantify the magnitude of fractionation gradients in an upcoming work, where we will also explore the observational implications of our work in more detail -- the main aim of the present paper was to present the newly developed model and to discuss some of the main results on a general level. On the theoretical side, extensions of the model to additional isotopes such as deuterium and $\rm ^{17}O$, $\rm ^{18}O$ would bring useful new information on the abundances of molecules fractionated in several species simultaneously, especially to constrain their usability as tracers of the shielded regions of prestellar cores.

\begin{acknowledgements}
We thank the anonymous referee for useful comments and suggestions. O.S. and P.C. gratefully acknowledge the funding by the Max Planck Society. L.C. acknowledges financial support through the Spanish grant PID2019-105552RB-C41 funded by MCIN/AEI/10.13039/501100011033. This work is supported by Chalmers Gender Initiative for Excellence (Genie).
\end{acknowledgements}

\bibliographystyle{aa}
\bibliography{47106}

\begin{thebibliography}{45}
\expandafter\ifx\csname natexlab\endcsname\relax\def\natexlab#1{#1}\fi

\bibitem[{{Adams} \& {Smith}(1981)}]{adams1981}
{Adams}, N.~G. \& {Smith}, D. 1981, \apjl, 247, L123

\bibitem[{{Altwegg} {et~al.}(2019){Altwegg}, {Balsiger}, \&
  {Fuselier}}]{Altwegg19}
{Altwegg}, K., {Balsiger}, H., \& {Fuselier}, S.~A. 2019, \araa, 57, 113

\bibitem[{{Anicich} {et~al.}(1977){Anicich}, {Huntress}, \&
  {Futrell}}]{anicich1977}
{Anicich}, V.~G., {Huntress}, W.~T., J., \& {Futrell}, J.~H. 1977, Chemical
  Physics Letters, 47, 488

\bibitem[{{Bizzocchi} {et~al.}(2013){Bizzocchi}, {Caselli}, {Leonardo}, \&
  {Dore}}]{Bizzocchi13}
{Bizzocchi}, L., {Caselli}, P., {Leonardo}, E., \& {Dore}, L. 2013, \aap, 555,
  A109

\bibitem[{{Caselli} \& {Ceccarelli}(2012)}]{Caselli12b}
{Caselli}, P. \& {Ceccarelli}, C. 2012, \aapr, 20, 56

\bibitem[{{Ceccarelli} {et~al.}(2014){Ceccarelli}, {Caselli},
  {Bockel{\'e}e-Morvan}, {Mousis}, {Pizzarello}, {Robert}, \&
  {Semenov}}]{Ceccarelli14}
{Ceccarelli}, C., {Caselli}, P., {Bockel{\'e}e-Morvan}, D., {et~al.} 2014, in
  Protostars and Planets VI, ed. H.~{Beuther}, R.~S. {Klessen}, C.~P.
  {Dullemond}, \& T.~{Henning}, 859--882

\bibitem[{{Colin} \& {Bernath}(2012)}]{Colin12}
{Colin}, R. \& {Bernath}, P.~F. 2012, Journal of Molecular Spectroscopy, 273,
  30

\bibitem[{{Colin} \& {Bernath}(2014)}]{Colin14}
{Colin}, R. \& {Bernath}, P.~F. 2014, Journal of Molecular Spectroscopy, 302,
  34

\bibitem[{{Colzi} {et~al.}(2019){Colzi}, {Fontani}, {Caselli}, {Leurini},
  {Bizzocchi}, \& {Quaia}}]{Colzi19}
{Colzi}, L., {Fontani}, F., {Caselli}, P., {et~al.} 2019, \mnras, 485, 5543

\bibitem[{{Colzi} {et~al.}(2018){Colzi}, {Fontani}, {Rivilla},
  {S{\'a}nchez-Monge}, {Testi}, {Beltr{\'a}n}, \& {Caselli}}]{Colzi18b}
{Colzi}, L., {Fontani}, F., {Rivilla}, V.~M., {et~al.} 2018, \mnras, 478, 3693

\bibitem[{{Colzi} {et~al.}(2020){Colzi}, {Sipil{\"a}}, {Roueff}, {Caselli}, \&
  {Fontani}}]{Colzi20}
{Colzi}, L., {Sipil{\"a}}, O., {Roueff}, E., {Caselli}, P., \& {Fontani}, F.
  2020, \aap, 640, A51 (C20)

\bibitem[{{Dislaire} {et~al.}(2012){Dislaire}, {Hily-Blant}, {Faure}, {Maret},
  {Bacmann}, \& {Pineau Des For{\^e}ts}}]{Dislaire12}
{Dislaire}, V., {Hily-Blant}, P., {Faure}, A., {et~al.} 2012, \aap, 537, A20

\bibitem[{{Draine} \& {Bertoldi}(1996)}]{Draine96}
{Draine}, B.~T. \& {Bertoldi}, F. 1996, \apj, 468, 269

\bibitem[{{Drozdovskaya} {et~al.}(2021){Drozdovskaya}, {Schroeder I}, {Rubin},
  {Altwegg}, {van Dishoeck}, {Kulterer}, {De Keyser}, {Fuselier}, \&
  {Combi}}]{Drozdovskaya21}
{Drozdovskaya}, M.~N., {Schroeder I}, I. R.~H.~G., {Rubin}, M., {et~al.} 2021,
  \mnras, 500, 4901

\bibitem[{{Faure} {et~al.}(2019){Faure}, {Hily-Blant}, {Rist}, {Pineau des
  For{\^e}ts}, {Matthews}, \& {Flower}}]{Faure19}
{Faure}, A., {Hily-Blant}, P., {Rist}, C., {et~al.} 2019, \mnras, 487, 3392

\bibitem[{{Furuya} \& {Aikawa}(2018)}]{Furuya18}
{Furuya}, K. \& {Aikawa}, Y. 2018, \apj, 857, 105

\bibitem[{{Grewal} {et~al.}(2021){Grewal}, {Dasgupta}, \& {Marty}}]{Grewal21}
{Grewal}, D.~S., {Dasgupta}, R., \& {Marty}, B. 2021, Nature Astronomy, 5, 356

\bibitem[{{Heays} {et~al.}(2014){Heays}, {Visser}, {Gredel}, {Ubachs}, {Lewis},
  {Gibson}, \& {van Dishoeck}}]{Heays14}
{Heays}, A.~N., {Visser}, R., {Gredel}, R., {et~al.} 2014, \aap, 562, A61

\bibitem[{{Hily-Blant} {et~al.}(2019){Hily-Blant}, {Magalhaes de Souza},
  {Kastner}, \& {Forveille}}]{Hily-Blant19}
{Hily-Blant}, P., {Magalhaes de Souza}, V., {Kastner}, J., \& {Forveille}, T.
  2019, \aap, 632, L12

\bibitem[{{Hily-Blant} {et~al.}(2020){Hily-Blant}, {Pineau des For{\^e}ts},
  {Faure}, \& {Flower}}]{Hily-Blant20}
{Hily-Blant}, P., {Pineau des For{\^e}ts}, G., {Faure}, A., \& {Flower}, D.~R.
  2020, \aap, 643, A76

\bibitem[{{Li} {et~al.}(2013){Li}, {Heays}, {Visser, R.}, {Ubachs, W.}, {Lewis,
  B. R.}, {Gibson, S. T.}, \& {van Dishoeck, E. F.}}]{Li13}
{Li}, X., {Heays}, A.~N., {Visser, R.}, {et~al.} 2013, A\&A, 555, A14

\bibitem[{{Loison} {et~al.}(2019){Loison}, {Wakelam}, {Gratier}, \&
  {Hickson}}]{loison2019}
{Loison}, J.-C., {Wakelam}, V., {Gratier}, P., \& {Hickson}, K.~M. 2019,
  \mnras, 484, 2747

\bibitem[{{Loison} {et~al.}(2020){Loison}, {Wakelam}, {Gratier}, \&
  {Hickson}}]{loison2020}
{Loison}, J.-C., {Wakelam}, V., {Gratier}, P., \& {Hickson}, K.~M. 2020,
  \mnras, 498, 4663

\bibitem[{{Lupi} {et~al.}(2021){Lupi}, {Bovino}, \& {Grassi}}]{Lupi21}
{Lupi}, A., {Bovino}, S., \& {Grassi}, T. 2021, \aap, 654, L6

\bibitem[{{Marty} {et~al.}(2011){Marty}, {Chaussidon}, {Wiens}, {Jurewicz}, \&
  {Burnett}}]{Marty11}
{Marty}, B., {Chaussidon}, M., {Wiens}, R.~C., {Jurewicz}, A.~J.~G., \&
  {Burnett}, D.~S. 2011, Science, 332, 1533

\bibitem[{{Mehnen} {et~al.}(2022){Mehnen}, {Suarez Martin}, {Roueff},
  {Hochlaf}, \& {Nyman}}]{Mehnen22}
{Mehnen}, B., {Suarez Martin}, I., {Roueff}, E., {Hochlaf}, M., \& {Nyman}, G.
  2022, \mnras, 517, 3126

\bibitem[{{Milam} {et~al.}(2005){Milam}, {Savage}, {Brewster}, {Ziurys}, \&
  {Wyckoff}}]{Milam05}
{Milam}, S.~N., {Savage}, C., {Brewster}, M.~A., {Ziurys}, L.~M., \& {Wyckoff},
  S. 2005, \apj, 634, 1126

\bibitem[{{Nomura} {et~al.}(2022){Nomura}, {Furuya}, {Cordiner}, {Charnley},
  {Alexander}, {Nixon}, {Guzman}, {Yurimoto}, {Tsukagoshi}, \&
  {Iino}}]{Nomura22}
{Nomura}, H., {Furuya}, K., {Cordiner}, M.~A., {et~al.} 2022, arXiv e-prints,
  arXiv:2203.10863

\bibitem[{{Ram} \& {Bernath}(2012)}]{Ram12}
{Ram}, R.~S. \& {Bernath}, P.~F. 2012, Journal of Molecular Spectroscopy, 274,
  22

\bibitem[{{Ram} {et~al.}(2010){Ram}, {Wallace}, \& {Bernath}}]{Ram10}
{Ram}, R.~S., {Wallace}, L., \& {Bernath}, P.~F. 2010, Journal of Molecular
  Spectroscopy, 263, 82

\bibitem[{{Redaelli} {et~al.}(2018){Redaelli}, {Bizzocchi}, {Caselli}, {Harju},
  {Chac{\'o}n-Tanarro}, {Leonardo}, \& {Dore}}]{Redaelli18}
{Redaelli}, E., {Bizzocchi}, L., {Caselli}, P., {et~al.} 2018, \aap, 617, A7

\bibitem[{{Redaelli} {et~al.}(2023){Redaelli}, {Bizzocchi}, {Caselli}, \&
  {Pineda}}]{Redaelli23}
{Redaelli}, E., {Bizzocchi}, L., {Caselli}, P., \& {Pineda}, J.~E. 2023, \aap,
  674, L8

\bibitem[{{Rocha} \& {Linnartz}(2021)}]{Rocha21}
{Rocha}, C.~M.~R. \& {Linnartz}, H. 2021, \aap, 647, A142

\bibitem[{{Roueff} {et~al.}(2015){Roueff}, {Loison}, \& {Hickson}}]{Roueff15}
{Roueff}, E., {Loison}, J.~C., \& {Hickson}, K.~M. 2015, \aap, 576, A99

\bibitem[{{Semenov} {et~al.}(2010){Semenov}, {Hersant}, {Wakelam}, {Dutrey},
  {Chapillon}, {Guilloteau}, {Henning}, {Launhardt}, {Pi{\'e}tu}, \&
  {Schreyer}}]{Semenov10}
{Semenov}, D., {Hersant}, F., {Wakelam}, V., {et~al.} 2010, \aap, 522, A42

\bibitem[{{Sipil{\"a}} {et~al.}(2015){Sipil{\"a}}, {Harju}, {Caselli}, \&
  {Schlemmer}}]{Sipila15b}
{Sipil{\"a}}, O., {Harju}, J., {Caselli}, P., \& {Schlemmer}, S. 2015, \aap,
  581, A122

\bibitem[{{Spezzano} {et~al.}(2016){Spezzano}, {Bizzocchi}, {Caselli}, {Harju},
  \& {Br{\"u}nken}}]{Spezzano16b}
{Spezzano}, S., {Bizzocchi}, L., {Caselli}, P., {Harju}, J., \& {Br{\"u}nken},
  S. 2016, \aap, 592, L11

\bibitem[{{Spezzano} {et~al.}(2022){Spezzano}, {Caselli}, {Sipil{\"a}}, \&
  {Bizzocchi}}]{Spezzano22b}
{Spezzano}, S., {Caselli}, P., {Sipil{\"a}}, O., \& {Bizzocchi}, L. 2022, \aap,
  664, L2

\bibitem[{{Visser} {et~al.}(2018){Visser}, {Bruderer}, {Cazzoletti},
  {Facchini}, {Heays}, \& {van Dishoeck}}]{Visser18}
{Visser}, R., {Bruderer}, S., {Cazzoletti}, P., {et~al.} 2018, \aap, 615, A75

\bibitem[{{Visser} {et~al.}(2009){Visser}, {van Dishoeck}, \&
  {Black}}]{Visser09}
{Visser}, R., {van Dishoeck}, E.~F., \& {Black}, J.~H. 2009, \aap, 503, 323

\bibitem[{{Wakelam} {et~al.}(2015){Wakelam}, {Loison}, {Herbst}, {Pavone},
  {Bergeat}, {B{\'e}roff}, {Chabot}, {Faure}, {Galli}, {Geppert}, {Gerlich},
  {Gratier}, {Harada}, {Hickson}, {Honvault}, {Klippenstein}, {Le Picard},
  {Nyman}, {Ruaud}, {Schlemmer}, {Sims}, {Talbi}, {Tennyson}, \&
  {Wester}}]{Wakelam15}
{Wakelam}, V., {Loison}, J.-C., {Herbst}, E., {et~al.} 2015, \apjs, 217, 20

\bibitem[{{Wampfler} {et~al.}(2014){Wampfler}, {J{\o}rgensen}, {Bizzarro}, \&
  {Bisschop}}]{Wampfler14}
{Wampfler}, S.~F., {J{\o}rgensen}, J.~K., {Bizzarro}, M., \& {Bisschop}, S.~E.
  2014, \aap, 572, A24

\bibitem[{{Watanabe} {et~al.}(2010){Watanabe}, {Kimura}, {Kouchi}, {Chigai},
  {Hama}, \& {Pirronello}}]{Watanabe10}
{Watanabe}, N., {Kimura}, Y., {Kouchi}, A., {et~al.} 2010, \apjl, 714, L233

\bibitem[{{Wirstr{\"o}m} \& {Charnley}(2018)}]{Wirstrom18}
{Wirstr{\"o}m}, E.~S. \& {Charnley}, S.~B. 2018, \mnras, 474, 3720

\bibitem[{{Wirstr{\"o}m} {et~al.}(2012){Wirstr{\"o}m}, {Charnley}, {Cordiner},
  \& {Milam}}]{Wirstrom12}
{Wirstr{\"o}m}, E.~S., {Charnley}, S.~B., {Cordiner}, M.~A., \& {Milam}, S.~N.
  2012, \apjl, 757, L11

\end{thebibliography}

\begin{appendix}


\section{Additional simulation results}\label{appendix:a}

\begin{figure*}
\centering
        \includegraphics[width=1.8\columnwidth]{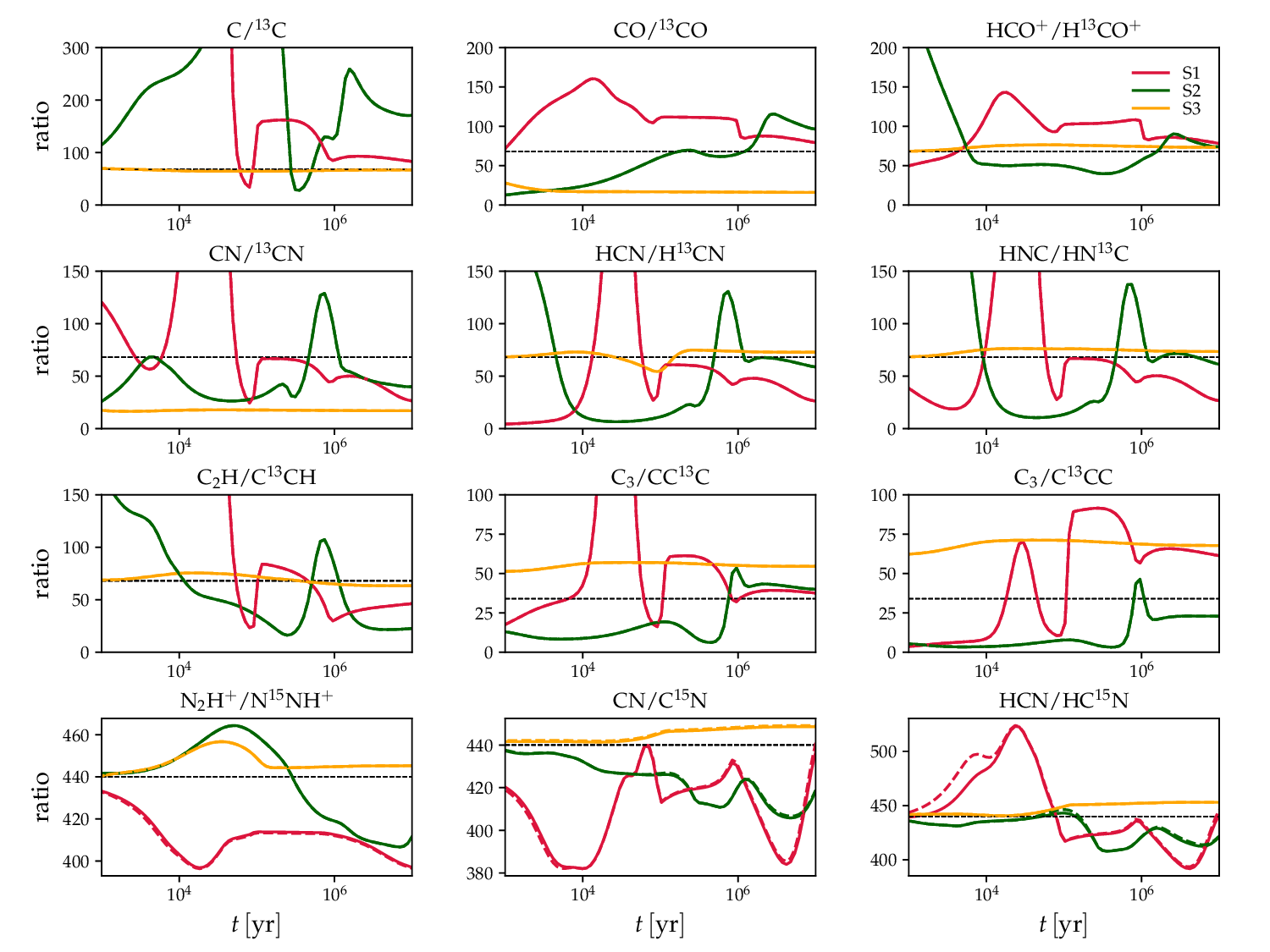}
    \caption{Selected $\cc/\cz$ and $\nn/\nz$ abundance ratios as a function of time in simulations S1~to~S3, labeled in the top right panel. The dashed black line in each panel indicates either the assumed elemental $\rm ^{14}N$/$\rm ^{15}N$ ratio (440) or the elemental $\rm ^{12}C$/$\rm ^{13}C$ ratio (68), except for $\rm C_3$ for which a ratio of 34 is used as our model does not distinguish between $\rm CC\cz$ and $\cz CC$. Solid and dashed colored lines represent simulations without and with the C+N exchange reactions (Table~\ref{tab:reaccolzi2}), respectively. In the case of the $\cc/\cz$ ratios, these lines overlap almost perfectly.}
    \label{fig:additionalResults}
\end{figure*}

Figure~\ref{fig:additionalResults} shows the effect of density and temperature changes on selected $\cc/\cz$ ratios, which we did not explore in Sect.\,\ref{s:results}. For completeness, we also include here plots for some species that were not included in Fig.\,\ref{fig:CfracComparison}. There is a great deal of variation in the $\cc/\cz$ ratios depending on the physical conditions, emphasizing the need for detailed models in interpreting the results of observations when using scaling factors is required. This could happen for example when one or several isotopic forms cannot be observed, and thus an abundance ratio cannot be consistently derived from observations. We note that the predictions of the lowest-density simulation (S3) are in this respect very different to the results presented in C20 for similar physical conditions (see especially their figures~7~and~B.1) -- the present model predicts a high level of fractionation for CO and CN, with $\cc/\cz$ ratios of only $\sim$10. This is because of the assumed visual extinction, which is 1\,mag in simulation~S3, but 10\,mag in C20. A low extinction translates to a high abundance of $\rm C^+$, enabling efficient fractionation through the exchange reactions that it is involved in (see Table~\ref{tab:reaccolzi}). We have confirmed via testing that if one increases the extinction such that photo-processes are no longer important, results similar to C20 are recovered also at low densities. Also, the lack of CO self-shielding in the fiducial simulation~S3 influences the degree of carbon fractionation (cf. Fig.\,\ref{fig:NC_selfShielding} in the main text).

The dependence of the $\rm N_2H^+/N\nz H^+$, $\rm CN/C\nz$, and $\rm HCN/HC\nz$ ratios on the physical conditions was already explored in Fig.\,\ref{fig:Nfrac} in the main text, but we include them in Fig.\,\ref{fig:additionalResults} in the context of demonstrating the impact of the C+N exchange reactions (Table~\ref{tab:reaccolzi2}) on singly fractionated carbon or nitrogen-bearing species. Evidently, the said reactions have a very marginal influence on the $\nn/\nz$ ratios, and negligible influence on the $\cc/\cz$ ratios.

\end{appendix}
\end{document}